\documentclass[aps,twocolumn,showpacs,amsmath,amssymb,floatfix]{revtex4-1}

\usepackage[dvips]{graphicx}
\usepackage{dcolumn}
\usepackage{bm}
\usepackage{amsmath}
\usepackage{amssymb}
\usepackage{ulem}
\usepackage{color} 
\usepackage[svgnames]{xcolor}
\usepackage[dvipdfmx,
colorlinks=true,
linkcolor=blue,
urlcolor=ForestGreen,
citecolor=magenta,
filecolor=blue,
pagecolor=blue,
linktocpage=true,
bookmarks=true,
bookmarksnumbered=true
]{hyperref}

\begin{document}

\title{Density matrix renormalization group study in energy space for a single-impurity Anderson model and an impurity quantum phase transition}

\author{Tomonori Shirakawa$^{1}$ and Seiji Yunoki$^{1,2,3}$}
\affiliation{
${}^1$Computational Quantum Matter Research Team, RIKEN Center for Emergent Matter Science (CEMS), Wako, Saitama 351-0198, Japan \\
${}^2$Computational Condensed Matter Physics Laboratory, RIKEN, Wako, Saitama 351-0198, Japan \\
${}^3$Computational Materials Science Research Team, RIKEN Advanced Institute for Computational Science (AICS), Kobe, Hyogo 650-0047, Japan 
}

\date{\today}

\begin{abstract} 
The density matrix renormalization group method is introduced in energy space to study 
Anderson impurity models.
The method allows for calculations in the thermodynamic limit and 
is advantageous for studying not only the dynamical properties but also 
the quantum entanglement of the ground-state at the vicinity of an impurity quantum phase transition.
This method is applied to obtain numerically exactly the ground-state phase diagram of the single-impurity 
Anderson model 
on the honeycomb lattice at half-filling. The calculation of local static quantities shows
that the phase diagram contains two distinct phases, the local moment (LM) phase and the asymmetric strong 
coupling (ASC) phase, but no Kondo screening phase. 
These results are supported by the local spin and charge excitation spectra, which exhibit 
qualitatively different behavior in these two phases 
and also reveal the existence of the valence fluctuating point at the phase boundary. 
For comparison, we also study the low-energy effective pseudogap Anderson model 
using the method introduced here. Although the high-energy excitations are obviously different, we find that 
the ground-state phase diagram and the asymptotically low-energy excitations are in good quantitative 
agreement with those for the single-impurity Anderson 
model on the honeycomb lattice, thus providing a quantitative justification for the previous studies 
based on low-energy approximate approaches. 
Furthermore, we find that the lowest entanglement level is doubly degenerate for the LM phase, whereas 
it is singlet for the ASC phase and is accidentally three fold degenerate at the valence fluctuating point. 
This should be contrasted with the degeneracy of the energy spectrum because the ground-state is 
found to be always singlet. 
Our results therefore clearly demonstrate that the low-lying entanglement spectrum can be used to determine 
with high accuracy the phase boundary of the impurity quantum phase transition. 
\end{abstract}


\maketitle

\section{Introduction}
Recent experiments have revealed that the introduction of adatoms 
induces the local magnetic moments in graphene~\cite{maccreary,nair}. 
One of the simplest models describing the magnetic impurity problem in graphene is an 
Anderson impurity 
coupled to the conduction electrons on the honeycomb lattice with the massless Dirac energy dispersion 
which results in 
the linear density of states ($\propto \left| \omega \right|$) around the Fermi level~\cite{neto}. 
This problem is known as the pseudogap Kondo problem~\cite{fritz1} and has been studied extensively 
for over a decade by the renormalization group analysis~\cite{vojta1,fritz2} and the numerical renormalization 
group (NRG) 
method~\cite{gonzalez-buxton,bulla1,bulla2,ingersent,kanao}. 
These previous studies have already found that, at zero temperature, there exit two phases, 
i.e., the local moment (LM) phase and the asymmetric strong coupling (ASC) phase, and the phase boundary 
corresponds to the valence fluctuating (VF) point~\cite{gonzalez-buxton}. 

These studies are, however, based on the low-energy effective pseudogap Anderson models~\cite{fritz1} 
and focused mostly on the two special cases in the large conduction band limit and in the strong 
coupling limit~\cite{gonzalez-buxton,bulla1,bulla2,ingersent}. 
Although the essential part of Kondo physics should be captured in these low-energy analyses, 
it is rather surprising that the phase diagram even for the simplest and most fundamental Anderson 
impurity model, not for the low-energy effective models, has not been established so far. 
The reason is simply because of lack of reliable numerical methods which can treat 
Anderson impurity models numerically exactly in two and three spatial dimensions. 
Establishing the numerically exact ground state phase diagram of the Anderson impurity 
model is also beneficial to the previous studies based on the low-energy effective models since 
it can provide a strong justification for their approximations. 

On the other hand, the entanglement spectrum~\cite{li} has attracted much attention recently 
in condensed matter physics for identifying topologically ordered phases~\cite{li,fidkowski,pollmann,yoshida,ejima}.
In these systems, the degeneracy of the low-lying entanglement spectrum is intimately related to the existence of the 
surface boundary states which are protected by bulk symmetries. 
Since the magnetic impurity problem can be regarded as a boundary problem in one dimension~\cite{affleck}, 
it is also valuable to explore its quantum entanglement aspects. 
In this context, the recent study of a ``spin-only'' version of a two-impurity Kondo model has shown 
that the gap of the entanglement spectrum can be regarded as an order parameter~\cite{bayat}. 
The entanglement spectrum is thus expected to be also used to quantify different quantum phases in magnetic 
impurity models including Anderson impurity models~\cite{vojta2}.

The main purposes of this paper are threefold. First, we introduce a numerical method 
which enables us to treat exactly general Anderson impurity models 
in any spatial dimension in the thermodynamic limit. 
Secondly, we demonstrate the method developed here by applying it to one of the simplest Anderson impurity 
models in two spatial dimensions and compare the results with those for the low-energy effective 
pseudogap Anderson model. Third,  
we explore the impurity quantum phase transition and the low-lying entanglement spectrum to uncover 
the degeneracy of the lowest entanglement level across the impurity quantum phase transition.

To this end, here we introduce the density matrix renormalization group (DMRG) 
method~\cite{white1,white2} for general Anderson impurity models represented in energy space. 
This method is applied to the single-impurity Anderson model on the honeycomb lattice at 
half-filling to determine precisely the ground-state phase diagram in a wide range of parameters, 
including an intermediate coupling region. The calculations of the local static quantities reveal that 
the phase diagram contains only two phases, i.e., the LM and ASC phases. 
The local spin and charge excitation spectra further support these results and also show 
the existence of the VF point at the phase boundary. 
We also study the low-energy effective pseudogap Anderson model and compare the 
ground-state phase diagram as well as the local spin and charge excitation spectra with those for the single-impurity 
Anderson model on the honeycomb lattice. 
Although the high-energy excitations are apparently different, the ground-state 
phase diagrams and the asymptotically low-energy excitations for these two models are found to be in 
excellent quantitative agreement. 
Moreover, we find that the degeneracy of the lowest entanglement level is different in the 
three different regions of the phase diagram, i.e., the LM and ASC phases and the VF point. 
This is in sharp contrast to the degeneracy of the ground state, which is found always singlet in all three regions. 
Although it has been pointed out that 
the sudden change of entanglement properties is not always related to quantum phase transition~\cite{chandran}, 
our results demonstrate that the low-lying entanglement spectrum can be used 
to determine the impurity quantum phase transition, at least, 
between the LM and ASC phases.

The rest of this paper is organized as follows. 
First, the single-impurity Anderson model on the honeycomb lattice and 
the corresponding low-energy effective pseudogap Anderson model are introduced 
in Sec.~\ref{sec:mm}. 
The DMRG method to solve general Anderson impurity models 
in energy space is also described in details in Sec.~\ref{sec:mm}. 
The numerical results for the single-impurity Anderson model on the honeycomb lattice 
are shown in Sec.~\ref{sec:results}. 
Based on the local static properties shown in Sec.~\ref{static}, the ground-state phase diagram 
is established in Sec.~\ref{sec:pd}. The local spin and charge excitation spectra are also 
calculated in Sec.~\ref{dynamical} to support the phase diagram. Furthermore, these results are 
compared with those for the low-energy effective pseudogap Anderson model in Sec.~\ref{sec:pga}. 
Finally, the low-lying entanglement spectrum and the entanglement entropy for the single-impurity 
Anderson model on the honeycomb lattice are discussed in Sec.~\ref{entanglement} 
before summarizing the paper in Sec.~\ref{summary}. The energy space description of the single-impurity 
Anderson model is discussed in Appendix~\ref{app:hr} and the further technical details of numerical 
calculations are provided in Appendix~\ref{app:nd}.

\section{\label{sec:mm}Models and Method}

In this section, we first introduce the single-impurity Anderson model on the honeycomb lattice.
Considering this model as an example, we describe in details the DMRG 
method in energy space, which can treat exactly general Anderson impurity models in any spatial dimensions. 
We also introduce the pseudogap Anderson model as a low-energy effective model for the 
single-impurity Anderson model on the honeycomb lattice.

\subsection{Single-impurity Anderson model}\label{sec:model}

\begin{figure}[tb]
\begin{center}
\includegraphics[width=\hsize]{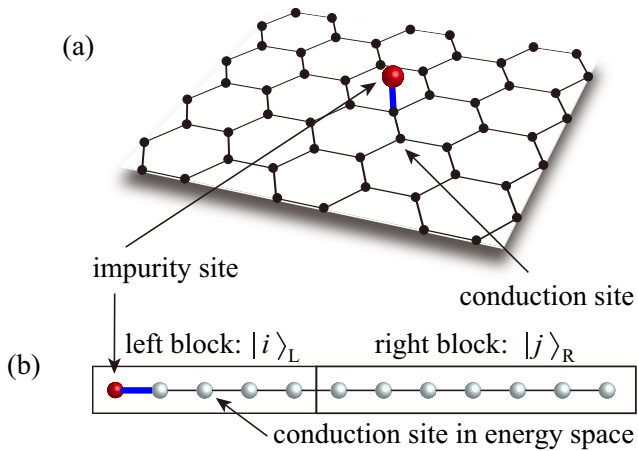}
\caption{(Color online) (a) Single-impurity Anderson model 
on the honeycomb lattice described by ${\mathcal H}_{\rm AIM}$ in Eq.~(\ref{eq:ham}). 
A red sphere and black dots represent the impurity site and the conduction sites, respectively. 
A blue line denotes the hybridization bond connecting the impurity site $i$ and the conduction site 
${\bf r}_0$ with the hybridization $V$ in Eq.~(\ref{eq:hv}), and black lines represent the lattice 
bonds connecting the nearest-neighbor conduction sites on the honeycomb lattice with 
the hopping $t$ in Eq.~(\ref{eq:hc}).
(b) Single-impurity Anderson model in energy space described by 
${\mathcal H}$ in Eq.~(\ref{eq:hamchain}). 
Note that 
although the impurity site (a red sphere at the left edge) is described by the same impurity Hamiltonian 
$\mathcal H_i$ in ${\mathcal H}_{\rm AIM}$ and 
is connected to only one of the conduction sites (gray spheres), 
the hybridization between the impurity site and the conduction site 
as well as the hopping between the conduction sites is generally different from those in (a).
In the DMRG calculations, 
the system is divided into two parts,  i.e., the left and right blocks, and the corresponding 
bases are denoted as $\left| i \right>_{\rm L}$ and $\left| j \right>_{\rm R}$.
 }
\label{fig:model}
\end{center}
\end{figure}

In order to be specific and also because it is one of the simplest models for 
the magnetic impurity problem in graphene, here we consider the single-impurity Anderson model 
on the honeycomb lattice 
[see Fig.~\ref{fig:model}(a)] defined by the following Hamiltonian: 
\begin{eqnarray}
\mathcal{H}_{\rm AIM} = \mathcal{H}_i + \mathcal{H}_c + \mathcal{H}_V, \label{eq:ham}
\end{eqnarray}
where the impurity Hamiltonian $\mathcal{H}_i $, including the on-site Coulomb interaction $U$ and 
the on-site potential $\varepsilon$ at the impurity site, is given as 
\begin{eqnarray}
\mathcal{H}_i = U n_{i,\uparrow} n_{i,\downarrow} - \varepsilon n_i, 
\end{eqnarray}
the conduction band with the nearest-neighbor hopping $t$ is described on the honeycomb lattice as 
\begin{eqnarray}
 \mathcal{H}_c = - t \sum_{\left< {\bf r},{\bf r}^{\prime}\right>} \sum_{\sigma = \uparrow, \downarrow}
( c_{{\bf r},\sigma}^{\dagger} c_{{\bf r}^{\prime},\sigma} + {\rm H.c.} ), 
\label{eq:hc}
\end{eqnarray}
and the hybridization between the impurity site and the conduction band is represented as 
\begin{eqnarray}
\mathcal{H}_V = V \sum_{\sigma = \uparrow,\downarrow}  ( c_{i,\sigma}^{\dagger} c_{{\bf r}_0,\sigma} 
+ {\rm H.c.} ). \label{eq:hv}
\end{eqnarray}
Here, $c_{{\bf r},\sigma}^{\dagger}$ ($c_{{\bf r},\sigma}$) is the creation (annihilation) operator of a 
conduction electron 
with spin $\sigma\, (= \uparrow,\downarrow)$ at site ${\bf r}$ on the honeycomb lattice and $c_{i,\sigma}^{\dagger}$ ($c_{i,\sigma}$)
is the creation (annihilation) operator of an electron at the impurity site $i$ with $n_{i,\sigma}= c_{i,\sigma}^{\dagger} c_{i,\sigma}$ 
and $n_i = n_{i, \uparrow} + n_{i, \downarrow}$. 
The sum in $\mathcal{H}_c$ denoted by $\langle {\bf r}, {\bf r}^{\prime} \rangle$ runs over nearest-neighbor 
pairs of conduction sites ${\bf r}$ and ${\bf r}^{\prime}$ on the honeycomb lattice. 
Notice here that the impurity site is connected to only one of the conduction sites at ${\bf r}_0$ 
through the hybridization $V$ in $\mathcal{H}_V$, as shown in Fig.~\ref{fig:model}(a). 
The electron density $n$ is set to be half-filled, i.e., $n=1$.

\subsection{DMRG method in energy space}\label{sec:dmrg}

We first describe the Hamiltonian $\mathcal{H}_{\rm AIM}$ in energy space. 
This can be done by noticing that the effective action of the impurity site for 
$\mathcal{H}_{\rm AIM}$ can be 
reproduced exactly by, e.g., the following Hamiltonian:
\begin{eqnarray}\label{eq:hw}
 \mathcal{H}_\omega &=& \mathcal{H}_i
+\sum_{\sigma = \uparrow, \downarrow}\int {\rm d}\omega \omega a^\dag_{\omega,\sigma}a_{\omega,\sigma} \nonumber \\
& +&V \sum_{\sigma = \uparrow,\downarrow} \int {\rm d}\omega  \sqrt{\rho (\omega)} \left( c^\dag_{i,\sigma}a_{\omega,\sigma}+{\rm H.c.}\right), 
\end{eqnarray}
where $a^\dag_{\omega,\sigma}$ ($a_{\omega,\sigma}$) is the creation (annihilation) 
operator of a conduction electron which represents the eigenstate of ${\cal H}_c$ with 
energy $\omega$ and spin $\sigma$, and the local density of state per spin for the conduction band 
is denoted as $\rho(\omega)$~\cite{gonzalez-buxton,bulla1}.
The equivalence between $\mathcal{H}_{\rm AIM}$ and $\mathcal{H}_{\omega}$ is shown 
in Appendix \ref{app:hr}.

\begin{figure}[tb]
\begin{center}
\includegraphics[width=\hsize]{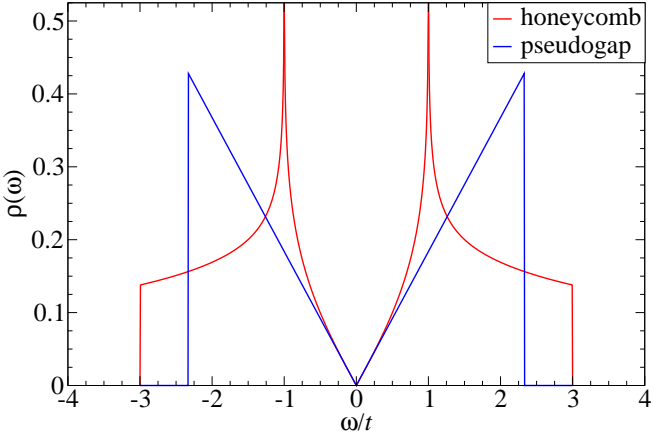}
\caption{(Color online) 
Local density of states $\rho(\omega)$ per spin for the conduction band of the 
single impurity Anderson model on the honeycomb lattice (red) and 
the pseudogap Anderson model (blue). 
Here, $W^2 = 4\sqrt{3}\pi t^2$ for the band width $W$ of the pseudogap Anderson model. 
The Fermi energy is located at $\omega=0$ for half filling . 
}
\label{fig:dos}
\end{center}
\end{figure}

Next, we discretize the energy $\omega$ with the logarithmic discretization scheme 
\begin{equation}
\omega_m^{\pm} = \pm \frac{W}{2} \Lambda^{-m}, 
\end{equation} 
where $W$ is the conduction band width ($W = 6t$ for the conduction band described by $\mathcal H_c$), 
$\Lambda\,(>1)$ is a parameter which sets a series of intervals 
in $\omega_m^{\pm}$'s with $m=0,1,\dots,M-1$~\cite{wilson}, and we set $\omega_{M}^{\pm} = 0$.
Defining 
a representative fermion operator $a^{\dagger}_{m,\pm,\sigma}$ ($m = 1, 2, \cdots, M$) 
for each energy interval between $\omega_{m-1}^{\pm}$ and $\omega_{m}^{\pm}$, 
the Hamiltonian $\mathcal{H}_\omega$ can now be expressed as
\begin{eqnarray}\label{eq:hr}
 \mathcal{H}_r &=& \mathcal{H}_i 
+\sum_{m=1}^{M} \sum_{\sigma = \uparrow,\downarrow} \left( \xi_{m}^+ a_{m,+,\sigma}^{\dagger} a_{m,+,\sigma} + \xi_{m}^{-} a_{m,-,\sigma}^{\dagger} a_{m,-,\sigma} \right) \nonumber \\
& + &\sum_{m=1}^{M}\sum_{\sigma = \uparrow, \downarrow} \left( \gamma_{m}^+ 
c_{i,\sigma}^{\dagger} a_{m,+,\sigma} + \gamma_m^- c_{i,\sigma}^{\dagger} a_{m,-,\sigma} + {\rm H.c.}\right),
\end{eqnarray}
where 
\begin{eqnarray}
 \gamma^{\pm}_m  = V \left[ \mp \int_{\pm\omega_{m}}^{\pm\omega_{m-1}} {\rm d}\omega \rho (\omega)\right]^{1/2}
\end{eqnarray}
and 
\begin{eqnarray}
 \xi_{m}^{\pm} = \frac{\int_{\pm\omega_{m}}^{\pm\omega_{m-1}} {\rm d}\omega \rho (\omega) \omega }
 {\int_{\pm\omega_{m}}^{\pm\omega_{m-1}} {\rm d}\omega \rho (\omega) } . 
\end{eqnarray}
This discretization scheme is similar to the one employed in the NRG method~\cite{bulla3}.
The local density of states $\rho(\omega)$ for the conduction band can be calculated with desired accuracy 
by employing the linear tetrahedron method, as shown in Fig.~\ref{fig:dos}~\cite{lehmann,jepsen}.

Since we can use any discretization scheme to discretize the energy $\omega$, we also introduce 
the constant discretization scheme 
\begin{equation}
\omega_m^{\pm} = \pm \frac{W}{2} \left(1 - \frac{m}{M}\right)
\end{equation} 
with $m=0,1,\dots,M$. The logarithmic discretization scheme has much denser energy 
meshes as $|\omega|$ approaches to zero, but has much less energy meshes for 
larger $|\omega|$ away from zero. 
Therefore, in order to treat correctly the band structure in high-energy scales, 
the logarithmic discretization scheme requires small $\Lambda$ close to 1, which is computationally 
demanding. 
In contrast, the constant discretization scheme distributes 
the energy meshes equally for all energy scales. We find that the logarithmic discretization scheme is 
suitable for the calculation of static quantities but the constant discretization scheme is 
better computationally to calculate dynamical quantities specially when the high-energy 
excitations are involved (see Appendix~\ref{app:nd} for more details). 
It is also shown in Appendix \ref{app:hr} that the effective action of the impurity site for $\mathcal H_r$ 
in both discretization schemes is exactly the same as the one for $\mathcal H_{\rm AIM}$.

Finally, we transform the Hamiltonian $\mathcal{H}_r$ into a one-dimensional form 
with no long-range hopping terms but 
keeping the interaction term local. For this purpose, we apply the Lanczos iteration~\cite{Lanczos} 
to the second and third terms 
in the right-hand side of ${\mathcal H}_r$ with choosing the impurity site as the initial Lanczos basis vector.
Introducing the vector representation of the electron creation and annihilation operators, e.g., 
\begin{eqnarray}
{\bf a}_{\sigma}^{\dagger} = 
&{}&(c_{i,\sigma}^{\dagger},a_{1,+,\sigma}^{\dagger}, a_{2,+,\sigma}^{\dagger},a_{3,+,\sigma}^{\dagger},\cdots, a^\dag_{M-1,+,\sigma}, a_{M,+,\sigma}^{\dagger}, \nonumber \\
&{}&a_{M,-,\sigma}^{\dagger}, a_{M-1,-,\sigma}^{\dagger},\cdots, a_{3,-,\sigma}^{\dagger}, a_{2,-,\sigma}^{\dagger}, a_{1,-,\sigma}^{\dagger}), 
\end{eqnarray}
$\mathcal{H}_{r}$ can be represented as
\begin{eqnarray}
\mathcal{H}_r = \mathcal{H}_i + \sum_{\sigma=\uparrow,\downarrow} {\bf a}_{\sigma}^{\dagger} \hat{H}_0 {\bf a}_{\sigma},
\end{eqnarray}
where $\hat{H}_0$ is a $(2M+1)\times(2M+1)$ matrix defined as 
\begin{eqnarray}
\hat{H}_0 = \left( 
\begin{array}{cccccccccc}
0 & \gamma_1^+ & \gamma_2^+ & \gamma_3^+ &  \cdots &\gamma_M^+ & \gamma_M^- &  \cdots &  \gamma_2^- & \gamma_1^-\\
\gamma_1^+ & \xi_1^+ & 0 & 0 & \cdots & 0 & 0 & \cdots & 0 & 0\\
\gamma_2^+ & 0 & \xi_2^+ & 0 & \cdots & 0 & 0 & \cdots & 0 & 0\\
\gamma_3^+ & 0 & 0 & \xi_3^+ & \cdots & 0 & 0 & \cdots & 0 & 0\\
\vdots & \vdots & \vdots & \vdots  & \ddots & \vdots & \vdots & \ddots & \vdots & \vdots\\
\gamma_M^+ & 0 & 0 & 0 &  \cdots & \xi_M^+ & 0 & \cdots & 0 & 0\\
\gamma_M^- & 0 & 0 & 0 &  \cdots & 0 & \xi_M^- & \cdots & 0 & 0\\
\vdots & \vdots & \vdots & \vdots  & \ddots & \vdots & \vdots & \ddots & \vdots & \vdots\\
\gamma_2^- & 0 & 0 & 0 &  \cdots & 0 & 0 & \cdots & \xi_2^- & 0\\
\gamma_1^- & 0 & 0 & 0 &  \cdots & 0 & 0 & \cdots & 0 & \xi_1^-\\
\end{array}
\right).
\end{eqnarray}
Notice here that although the interaction term at the impurity site is described by the same 
impurity Hamiltonian ${\cal H}_i$, $\mathcal H_r$ introduces long-range hopping terms as the impurity site is 
hybridized with all conduction sites in energy space. This is 
usually problematic for DMRG calculations. As shown below, this can be completely alleviated by the 
Lanczos basis transformation without introducing additional long-range interaction terms for $\mathcal H_i$.

Taking as the initial Lanczos basis the $(2M+1)$-dimensional column unit vector ${\bf p}_1$ with the $k$th element 
\begin{equation}
({\bf p}_1)_k = \delta_{k,1} \quad(k=1,2,\cdots,2M+1),  
\end{equation}
we can generate the Lanczos basis via the three-time recurrences
\begin{eqnarray}
t_{l} {\bf p}_{l+1} = \hat{H}_0 {\bf p}_l - \varepsilon_l {\bf p}_l - t_{l-1} {\bf p}_{l-1} 
\label{eq:lc}
\end{eqnarray}
for $l=1,2,\cdots, L-1$ with $t_0=0$ and ${\bf p}_0 = 0$, where 
\begin{equation}
\varepsilon_l = {\bf p}_l^T{\hat H}_0{\bf p}_l
\end{equation}
and 
\begin{equation}
t_l=\left|{\hat H}_0{\bf p}_l-\varepsilon_l {\bf p}_l - t_{l-1} {\bf p}_{l-1}\right|, 
\end{equation}
i.e., the norm of the $(2M+1)$-dimensional column vector in the right-hand side of 
Eq.~(\ref{eq:lc}).

Using these Lanczos bases $\hat{P} = ({\bf p}_1, {\bf p}_2, \cdots,{\bf p}_L)$, 
$\hat{H}_0$ can be transformed into the following tridiagonal matrix: 
\begin{eqnarray}
\hat{H}_0^{\prime} = \left( 
\begin{array}{ccccccc}
\varepsilon_1 & t_1 & 0 & 0 & \cdots & 0 & 0\\
t_1 & \varepsilon_2 & t_2 & 0 & \cdots& 0 & 0 \\
0 & t_2 & \varepsilon_3 & t_3 & \cdots & 0 & 0\\
0 & 0 & t_3 & \varepsilon_4 & \cdots & 0 & 0\\
\vdots & \vdots & \vdots & \vdots & \ddots& \vdots & \vdots\\
0 & 0 & 0 & 0 & \cdots &  \varepsilon_{L-1}& t_{l-1}\\
0 & 0 & 0 & 0 & \cdots &  t_{l-1}& \varepsilon_{L}\\
\end{array}
\right) \label{eq:tridiagonalform}
\end{eqnarray}
and accordingly the electron creations operators ${\bf a}^\dag_\sigma$ are transformed into the 
new operators ${\bf f}^\dag_\sigma$ as 
\begin{equation}
{\bf f}_{\sigma}^{\dagger} = ( f_{1,\sigma}^{\dagger}, f_{2,\sigma}^{\dagger}, \cdots, f_{L,\sigma}^{\dagger} ) = {\bf a}_{\sigma}^\dag \hat{P}
\end{equation}
with $f_{1,\sigma}^\dag = c_{i,\sigma}^\dag$. Since $({\bf p}_l)_k=({\hat P})_{k,l}$ and 
${\bf p}_l \cdot {\bf p}_{l^{\prime}} = \delta_{l,l^{\prime}}$, 
one can easily show that $({\hat P}^T{\hat P})_{l,l'}= \delta_{l,l^{\prime}}$. 
Therefore, the new operators ${\bf f}^\dag_\sigma$ and ${\bf f}_\sigma$ satisfy the fermion anticommutation relations, 
e.g., $\{f_{l,\sigma},f_{l',\sigma'}^\dag\}=\delta_{l,l^{\prime}}\delta_{\sigma,\sigma'}$. 
Note also that $\varepsilon_1=0$ as $({\hat H}_0)_{1,1}=0$.

The resulting Hamiltonian after this basis transformation is 
\begin{eqnarray}
\mathcal{H} &=& \mathcal{H}_i 
+ \sum_{\sigma = \uparrow,\downarrow} \sum_{l=1}^L \varepsilon_l f_{l,\sigma}^{\dagger} f_{l,\sigma} \nonumber \\
& +& \sum_{\sigma = \uparrow,\downarrow} \sum_{l=1}^{L-1} t_l \left (f_{l+1,\sigma}^{\dagger} f_{l,\sigma} + {\rm H.c.} \right).  
\label{eq:hamchain}
\end{eqnarray}
Notice first that the impurity site is described by the same local Hamiltonian $\mathcal H_i$ as in 
$\mathcal H_{\rm AIM}$ and therefore the interaction term remains local. 
On the other hand, the hopping terms are now all short ranged with only nearest-neighbor hopping $t_l$. 
Therefore, the model described by $\mathcal{H}$ is a simple one-dimensional system of $L$ sites, 
as schematically shown in Fig.~\ref{fig:model}(b), 
and can be best treated by the DMRG method~\cite{white1,white2,garcia,peters}.

Four remarks are in order regarding the method introduced here. 
First, as it is well known in the standard Lanczos method~\cite{Lanczos}, the transformation 
from ${\cal H}_r$ to $\cal H$ is exact only when $L=2M+1$, assuming that ${\bf p}_1$ is not 
contained in an invariant subspace and thus the Lanczos iteration is not terminated before generating 
${\bf p}_{2M+1}$. Only in this case, $({\hat P}{\hat P}^T)_{k,k'}=\delta_{k,k'}$ and thus 
\begin{eqnarray}
 {\bf a}_{\sigma}^{\dagger} \hat{H}_0 {\bf a}_{\sigma} = 
{\bf a}_{\sigma}^{\dagger} \hat{P} \hat{P}^T \hat{H}_0 \hat{P} \hat{P}^T {\bf a}_{\sigma} 
=  {\bf f}_{\sigma}^{\dagger} \hat{H}_0^{\prime} {\bf f}_{\sigma} 
\end{eqnarray}
as $\hat P$ is a $(2M+1)\times(2M+1)$ orthogonal matrix.

Second, it is apparent from the 
construction that this method is for calculations in the thermodynamic limit. 
$M$ and also $\Lambda$ in the logarithmic discretization scheme determine the energy 
resolution as well as the model parameters $\varepsilon_l$ and $t_l$ in $\cal H$. 
Therefore, these quantities $M$ and $\Lambda$ control the accuracy of $\cal H$ with respect to 
${\cal H}_{\rm AIM}$ 
in the thermodynamic limit. In principle, 
the logarithmic discretization scheme 
becomes exact when $\Lambda\to1+0^+\,(=1^+)$ and $M\to\infty$, where $0^+$ is 
positive infinitesimal. Similarly, the constant discretization scheme becomes exact when $M\to\infty$. 
However, as discussed in Appendix \ref{app:nd}, we find that reasonably large $M$ and $L$ 
(but $L\leq2M+1$) can well represent the thermodynamic limit  
(provided that $\Lambda$ is sufficiently small for the logarithmic discretization scheme).

Third, the single-impurity Anderson model in two spatial dimensions is considered here. 
This is only to simplify the explanation of the method introduced here. 
The extension of the method to more general cases such as a two-impurity Anderson model 
and a multiorbital many-impurity Anderson model is rather 
straightforward by using the block Lanczos technique~\cite{shirakawa}. 
Indeed, the similar transformation for Anderson impurity models in real space 
and its extension to multiorbital systems are found in Refs.~\cite{shirakawa,busser,allerdt}. 
The extension to three-dimensional systems is also straightforward.

Fourth, the method is similar to the NRG approach~\cite{bulla3} in that both 
treat Anderson impurity models represented in energy space. 
Indeed, the logarithmic discretization scheme is employed in the NRG method, 
where $M$ corresponds to the number of renormalization iterations~\cite{bulla3}.
However, these two approaches are conceptually different because the DMRG method optimizes 
the wave function based on the largest eigenstates of the reduced density matrix for that wave function, 
while the NRG method constructs the low-energy effective Hamiltonian based on the lowest 
eigenstates of the Hamiltonian. 
Note also that the method introduced here can take into account the band structure effect accurately 
over large energy scales, 
in contrast to the low-energy approximate approaches. This is obviously important when high-energy 
excitations are discussed.

\subsection{\label{sec:mpga}Pseudogap Anderson model}

It is valuable to compare the results for the single-impurity Anderson model on the honeycomb lattice 
with those for the corresponding pseudogap Anderson model. The pseudogap Anderson model is described 
by Hamiltonian $\mathcal H_\omega$ or $\mathcal H_r$ 
with the following local density of states $\rho_{\rm PGA}(\omega)$ per spin for the 
conduction band (see Fig.~\ref{fig:dos}): 
\begin{equation}
\rho_{\rm PGA}(\omega) = \left\{
\begin{array}{cc}
4|\omega|/W^2 & {\rm for}\ |\omega| \le W/2, \\
0 & {\rm for}\ |\omega|>W/2. \\
\end{array}
\right..  
\end{equation}
This local density of states is directly used for 
$\rho(\omega)$ in Eqs.~(\ref{eq:hw}) and (\ref{eq:hr}) to construct $\cal H$, which then can 
be solved by the DMRG method, as described above. 
The asymptotic behavior of $\rho (\omega)$ for the conduction band 
on the honeycomb lattice described by $\mathcal H_c$ is 
\begin{eqnarray}
\rho (\omega ) \sim \frac{1}{\sqrt{3} \pi t^2} \left| \omega \right|
\end{eqnarray}
for $\omega$ around zero~\cite{neto}. 
Therefore, by setting the band width $W$ for the pseudogap Anderson model as $W^2 = 4\sqrt{3} \pi t^2$, 
$\rho_{\rm PGA}(\omega)$ can reproduce exactly the asymptotic behavior of $\rho (\omega)$ for the conduction 
band on the honeycomb lattice, 
as shown in Fig.~\ref{fig:dos}.

The pseudogap Anderson model and its strong coupling counterpart, i.e., the pseudogap Kondo 
model, have been studied extensively using the analytical and NRG methods~\cite{fritz1,vojta1,fritz2,gonzalez-buxton,bulla1,bulla2,ingersent,kanao}. 
The ground-state phase diagram of the pseudogap Anderson model has two distinct phases, 
the LM phase and the ASC phase, and the VF point at the phase boundary~\cite{gonzalez-buxton}. 
In the LM phase, a free local moment at the impurity site survives  
without Kondo screening even at zero temperature. 
The fixed point of this phase can be characterized as 
$(\varepsilon, U, V)\to(\varepsilon^{\ast}, U^{\ast}, V^{\ast}) = (\varepsilon^{\ast}, 2\varepsilon^{\ast}, 0)$ 
with $\varepsilon^{\ast} = \infty$~\cite{gonzalez-buxton}, 
i.e., the free local moment being decoupled from the conduction band. 
In the ASC phase, 
the fixed point is characterized as $(\varepsilon^{\ast}, U^{\ast}, V^{\ast}) = (\infty,0,0)$ for $\varepsilon > U/2$, 
and $(\varepsilon^{\ast}, U^{\ast}, V^{\ast}) = (-\infty, 0, 0)$ for $\varepsilon < U/2$~\cite{gonzalez-buxton}. 
Since the two electrons (two holes) occupy the impurity site for $\varepsilon > U/2$ ($\varepsilon < U/2$), 
no local moment is formed. 
The fixed point for the VF point is characterized as 
$(\varepsilon^{\ast}, U^{\ast}, V^{\ast}) = (0, \infty, 0)$ for $\varepsilon < U/2$ 
and $(\varepsilon^{\ast}, U^{\ast}, V^{\ast}) = (\varepsilon^{\ast}, \varepsilon^{\ast}, 0)$ 
with $\varepsilon^{\ast} = \infty$ for $\varepsilon > U/2$~\cite{gonzalez-buxton}. 
Since the impurity site is decoupled from the conduction band, the ground state 
is threefold degenerate due to three different local states at the impurity site, 
i.e., the empty state and the singly occupied states with up or down electron for $\varepsilon < U/2$.

\section{\label{sec:results}Results}

We first summarize briefly the conditions employed for 
the numerical calculations before discussing the numerical results for the single-impurity Anderson 
model on the honeycomb lattice. 
These results are then compared with those for the pseudogap Anderson model. 
We also examine the entanglement spectrum for the ground state of the single-impurity Anderson model 
on the honeycomb lattice to characterize the impurity quantum phase transition.

\subsection{\label{sec:numericaldetail}Numerical details}

We set that $L=M$ (even) in Eq.~(\ref{eq:hamchain}) throughout the calculations discussed here, 
unless otherwise stated. 
As already explained in Sec.~\ref{sec:dmrg}, $M$ (and also $\Lambda$ when the logarithmic discretization 
scheme is used) controls the energy resolution and $L$ can be taken to be 
up to $2M+1$. Therefore, in principle, one should take the infinite limit of $M$ and $L$ along with 
$\Lambda\to1^+$ for the logarithmic discretization scheme. 
However, we find that the quantities studied here are well converged for sufficiently large but finite values of 
$M$ and $L$ with keeping a fixed ratio of $L/M$, at least, when the logarithmic discretization scheme with 
$\Lambda$ as small as 1.15 is used (see Appendix~\ref{app:nd}). 
Therefore, we take $L\, (=M)$ up to $128$ with the $z$ component of total spin $S_z=0$ and 
keep $m_{\rm D} \sim 32 L$ largest eigenstates of the reduced density matrix 
in the DMRG calculations.

When the logarithmic discretization scheme is used, the discarded weights 
are found to be significantly small, typically of the order $10^{-13}$--$10^{-11}$. 
The corresponding error of the ground-state energy is as small as $\sim 10^{-8} t$, 
which is even smaller than the smallest 
level spacing of the eigenvalues for $\hat{H}^{\prime}_0$ in Eq.~(\ref{eq:tridiagonalform}) with $L=M$. 
When the constant discretization scheme is used, the discarded weights are as small as $10^{-9}$ 
and the corresponding error of the ground-state energy is about $10^{-7} t$.
We employ the logarithmic discretization scheme to calculate the static quantities as well as 
the entanglement spectrum and the constant discretization scheme to calculate 
the full excitation spectra (see Appendix \ref{app:nd}). 
However, to extrapolate the static limit of the dynamical quantity, we use the logarithmic discretization scheme 
with $m_{\rm D}$ up to $48L$. 
It should be also noted that the ground state is found to be always singlet as long as $L$ is even and finite.

\subsection{\label{static}Local static quantities}

Let us first calculate the local density per spin at the impurity site 
\begin{eqnarray}
\bar{n}_{i\sigma} = \left< \psi_0 \right| n_{i\sigma} \left| \psi_0 \right> 
\end{eqnarray}
and the total spin at the impurity site 
\begin{eqnarray}
\bar{S}_i = \left< \psi_0 \right| \vec{S}_i \cdot \vec{S}_i \left| \psi_0 \right>, 
\label{eq:si}
\end{eqnarray}
where $\left| \psi_0 \right>$ is the ground state and the spin operator $\vec{S}_i$ 
at the impurity site is given as 
\begin{eqnarray}
\vec{S}_i = \frac{1}{2} \sum_{\sigma_1,\sigma_2} {c}_{i,\sigma_1}^{\dagger} \vec{\tau}_{\sigma_1,\sigma_2} {c}_{i,\sigma_2}
\end{eqnarray} 
with $\vec{\tau}=(\tau_x,\tau_y,\tau_z)$ being Pauli matrices. Note that 
$\bar{n}_{i\uparrow} =\bar{n}_{i\downarrow}$ because of ${\rm SU}(2)$ symmetry of 
$\cal H$ and $\bar{S}_i \le0.75$.

\begin{figure}[htbp]
\begin{center}
\includegraphics[width=\hsize]{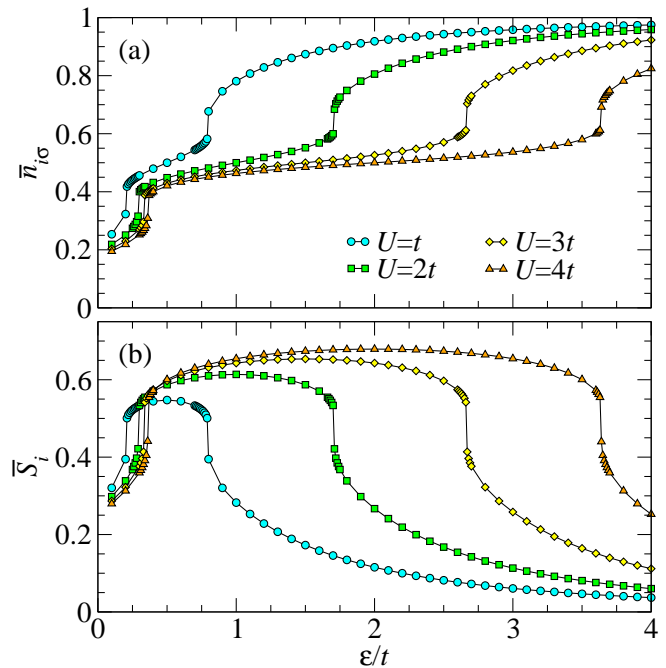}
\caption{(Color online) (a) Local density $\bar{n}_{i\sigma}$ and (b) total spin $\bar{S}_{i}$ at the impurity site 
for the single-impurity Anderson model on the honeycomb lattice 
as a function of $\varepsilon$ with $V=t$ and various $U$ indicated in (a). 
The logarithmic discretization scheme with $\Lambda = 1.15$ and $M=128$ is used.
}
\label{fig:ns}
\end{center}
\end{figure}

\begin{figure}[htbp]
\begin{center}
\includegraphics[width=\hsize]{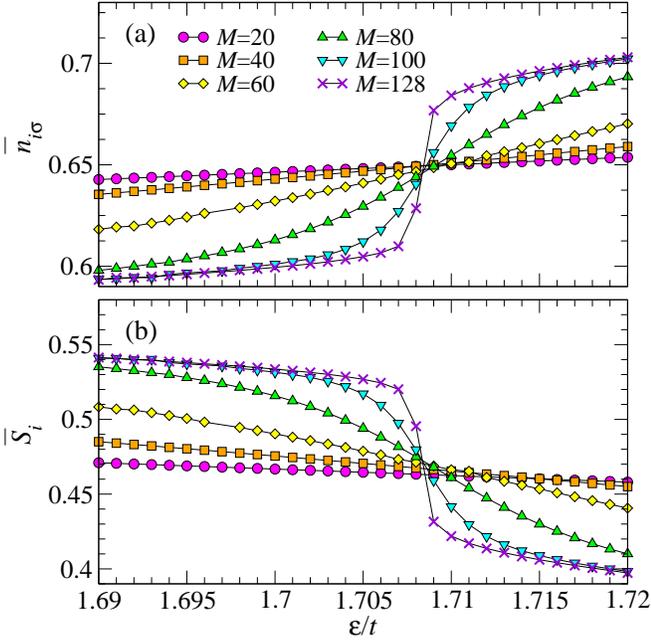}
\caption{(Color online) 
Same as in Fig.~\ref{fig:ns} but with $U=2t$ and a finer grid size of $\varepsilon$ for different values of 
$M$ indicated in (a). 
}
\label{fig:nsmdep}
\end{center}
\end{figure}

As shown in Fig.~\ref{fig:ns}, we find that 
these quantities change discontinuously at two distinct values of $\varepsilon$ for given $U$ and $V$. 
It should be emphasized, however, that these deceptively discontinuous changes are simply due to a 
finite grid size of $\varepsilon$ used in the figures, but not due to the level 
crossing of two different states as often found in finite-size calculations. 
Indeed, we find in Fig.~\ref{fig:nsmdep} that these quantities vary smoothly with $\varepsilon$ 
when a much smaller grid size of $\varepsilon$ is used. 
We also find in Fig.~\ref{fig:nsmdep} that the change of these quantities 
becomes sharper and steeper with increasing $M$. Therefore, we expect that it becomes truly 
discontinuous only when we take the limit of $M\to\infty$.

Nevertheless, the steep changes of these quantities imply that there are three phases for a given $U$, 
a low-density phase ($\bar{n}_{i\sigma} \alt 0.3$) for small $\varepsilon$, 
an intermediate density phase ($\bar{n}_{i\sigma} \sim 0.5$) which includes the particle-hole 
symmetric limit with $\varepsilon = U/2$, 
and a high-density phase ($\bar{n}_{i\sigma} \agt 0.7$) for large $\varepsilon$. 
While the total spin $\bar{S}_i$ is suppressed in the low- and high-density phases, it is enhanced in 
the intermediate density phase.  
It should be noticed here that the low- and high-density phases are related under the particle-hole 
transformation. 
This is because the model with a parameter set $(\varepsilon, U, V)$ in particle picture can be 
transformed into the same model 
with $(U - \varepsilon, U, V)$ in hole picture. Indeed, our results satisfy that $\bar{n}_{i\sigma}$ 
($\bar{S}_i$) at the impurity potential $\varepsilon$ is exactly the same as $1-\bar{n}_{i\sigma}$ ($\bar{S}_i$) at 
the impurity potential $U - \varepsilon$ for given $U$ and $V$. 
Together with the results of the spin and charge excitation spectra shown in Sec.~\ref{sec:pd}, 
we identify the intermediate density phase as the LM 
phase where the local moment is formed at the impurity site, and the low- and high-density phases as 
the ASC phase where essentially 
two holes or electrons occupy the impurity site with no local moment formed.

\subsection{\label{sec:pd}Ground-state phase diagram}

\begin{figure}[htbp]
\includegraphics[width=0.9\hsize]{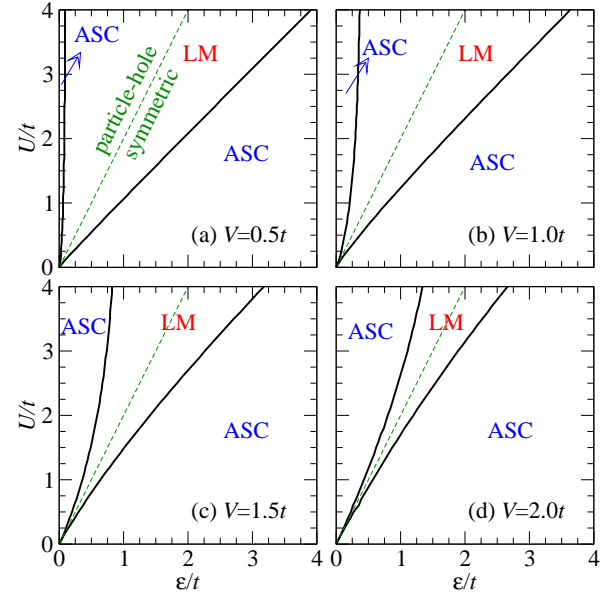}
\caption{(Color online) 
Ground-state phase diagrams for the single-impurity Anderson model 
on the honeycomb lattice in a wide range of parameters, 
$\varepsilon$ and $U$, with four different values of $V$ indicated in the figures. 
LM (ASC) stands for the local moment 
(asymmetric strong coupling) phase. The particle-hole symmetric line 
$\varepsilon  = U/2$ is indicated by green dashed lines. 
The phase boundaries are determined by the calculations of the static quantities shown in Fig.~\ref{fig:ns}. 
}
\label{fig:pd}
\end{figure}

Systematically calculating $\bar{n}_{i\sigma}$ and $\bar{S}_i$ for different values of $V$, 
we obtain the ground-state phase diagram for the single-impurity Anderson model 
on the honeycomb lattice in a wide range of parameters, 
$\varepsilon$ and $U$. 
As shown in Fig.~\ref{fig:pd}, the LM phase appears around the particle-hole symmetric limit with 
$\varepsilon=U/2$, where $\bar{n}_{i\sigma}$ is exactly $1/2$. The region of this phase is found 
to decrease with increasing $V$. 
This is easily understood by considering that the increase of $V$ enhances the bonding between 
the impurity site and the 
conduction site ${{\bf r}_0}$, which then leads to the formation of the bond singlet state. 
Eventually, the LM phase exists only along the particle-hole symmetric line in the limit of $V\to\infty$.  
Our results in Fig.~\ref{fig:pd} also imply the absence of Kondo screening phase. 
This can be understood simply as the consequence of the characteristic 
density of states of the conduction band since the Kondo temperature 
$T_{\rm K} \sim \sqrt{U V^2 \rho (0)/2} {\rm exp}\left[-\pi \left| (U - \varepsilon ) \varepsilon \right| / 4 U V^2 \rho(0)\right]$~\cite{hewson}. 
The phase diagram is therefore in good qualitative agreement with that obtained by the low-energy 
approximate approaches~\cite{fritz2,gonzalez-buxton}. 
In fact, as discussed below in Sec.~\ref{sec:pga}, the phase diagram is found to be quantitatively 
compared with that for the pseudogap Anderson model.

\subsection{\label{dynamical}Dynamical quantities}

Next, we calculate the dynamical quantities to support the assignment of different phases 
found in the phase diagram. 
The dynamical quantities studied here are the spin excitation spectrum at impurity site 
\begin{eqnarray}
\chi_{\rm s} (\omega) = - \frac{1}{\pi} {\rm Im}
\left< \psi_0 \right| S_i^z \frac{1}{\omega + {\rm i} \eta - \mathcal{H} + E_0} S_i^z 
\left| \psi_0 \right> 
\end{eqnarray}
and the charge excitation spectrum at impurity site 
\begin{eqnarray}
\chi_{\rm c} (\omega) = -\frac{1}{\pi} {\rm Im} 
\left< \psi_0 \right| n_i \frac{1}{\omega + {\rm i} \eta - \mathcal{H} + E_0} n_i
\left| \psi_0 \right>, 
\end{eqnarray}
where $E_0$ is the ground-state energy and $\eta$ is a broadening factor. 
We calculate these quantities using the dynamical DMRG method~\cite{jeckelmann}.

\begin{figure}[tbhp]
\includegraphics[width=\hsize]{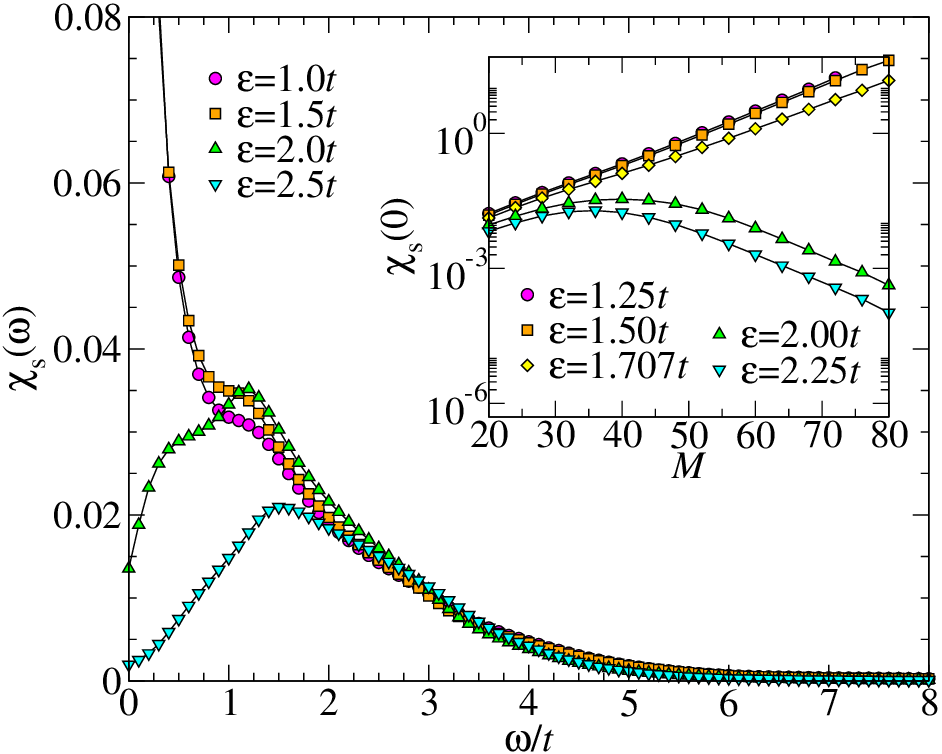}
\caption{(Color online) 
Spin excitation spectra $\chi_{\rm s}(\omega)$ at the impurity site 
for the single-impurity Anderson model on the honeycomb lattice with 
$V=t$ and $U=2t$. The constant discretization scheme with $M=100$ is used and 
a broadening factor of $\eta = 0.2t$ is set. 
The inset shows the $M$ dependence of $\chi_{\rm s}(0)$ calculated using the logarithmic 
discretization scheme with $\Lambda = 1.15$ and 
$\eta = 25W \Lambda^{-M}$. 
}
\label{fig:xs_hc}
\end{figure}

Figure~\ref{fig:xs_hc} shows the spin excitation spectra $\chi_{\rm s} (\omega)$ 
for $V = t$ and $U = 2t$. In this case, the transition occurs 
at $\varepsilon = \varepsilon_c \sim 1.707 t$--$1.708t$. 
As seen in Fig.~\ref{fig:xs_hc}, $\chi_{\rm s} (\omega)$ in the LM phase for $\varepsilon<\varepsilon_c$ 
increases as $\omega\to0$. 
In contrast, $\chi_{\rm s} (\omega)$ for $\omega \to0$ is suppressed in the ASC phase for 
$\varepsilon>\varepsilon_c$. 
The $M$ dependence of $\chi_{\rm s} (0)$ shown in the inset of Fig.~\ref{fig:xs_hc} 
reveals that $\chi_{\rm s}(0)$ in the LM (ASC) phase increases (decreases) 
exponentially with $M$ for large $M$, i.e., 
\begin{equation}
\chi_{\rm s}(0) \propto \exp\left( \tau_{\rm s}M \right) \label{eq:chisexp}
\end{equation} 
with the same $\tau_{\rm s}$ in each phase, independent of values of $\varepsilon$, 
as summarized in Table~\ref{tab:exp_hc}. 
Therefore, we can safely conclude that 
\begin{equation}
\lim_{M\to\infty}\chi_{\rm s}(0) \to \infty  
\end{equation}
in the LM phase and 
\begin{equation}
\lim_{M\to\infty}\chi_{\rm s}(0) \to 0 
\end{equation}
in the ASC phase. 
The divergent behavior of $\chi_{\rm s}(0)$ implies the presence of free local moment 
in the LM phases, 
while $\chi_{\rm s}(0)=0$ indicates no local moment at the impurity site in the ASC phase.

\begin{table}[htbp]
\caption{The diverging or decaying factors $\tau_{\rm s}$ and $\tau_{\rm c}$ of the local spin 
and charge excitation spectra at $\omega=0$, $\chi_{\rm s}(0)\propto\exp(\tau_{\rm s}M)$, 
and $\chi_{\rm c}(0)\propto\exp(\tau_{\rm c}M)$, respectively, 
for large $M$ in the three different regions of the phase diagram, i.e., the LM and ASC phases and the VF point, 
estimated from the results shown in Figs.~\ref{fig:xs_hc} and \ref{fig:xc_hc}. }\label{tab:exp_hc}
\begin{tabular}{c|ccc}
\hline \hline
   {}   &  LM phase & ASC phase & VF point \\
      \hline
$\tau_{\rm s}$  & $0.140$ & $-0.140$ & $0.123$ \\
$\tau_{\rm c}$  & $-0.143$ & $-0.142$ & $0.082$ \\
\hline \hline
\end{tabular}
\end{table}

\begin{figure}[tbhp]
\includegraphics[width=\hsize]{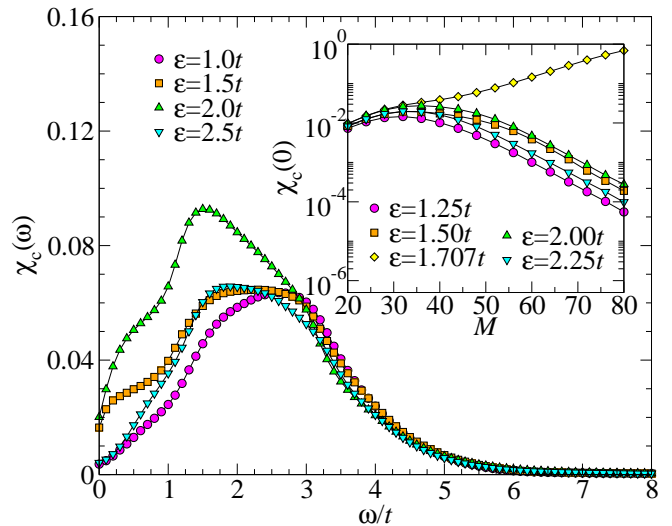}
\caption{(Color online) 
Charge excitation spectra $\chi_{\rm c}(\omega)$ at the impurity site 
for the single-impurity Anderson model on the honeycomb lattice with 
$V=t$ and $U=2t$. The constant discretization scheme with $M=100$ is used and 
a broadening factor of $\eta = 0.2t$ is set. 
The inset shows the $M$ dependence of $\chi_{\rm c}(0)$ calculated using the logarithmic 
discretization scheme with $\Lambda = 1.15$ and 
$\eta = 25W \Lambda^{-M}$. 
}
\label{fig:xc_hc}
\end{figure}

The charge excitation spectra $\chi_{\rm c} (\omega)$ are shown in Fig.~\ref{fig:xc_hc}. 
In contrast to $\chi_{\rm s}(\omega)$, we find that $\chi_{\rm c} (\omega) \to 0$ for $\omega\to0$ in 
both LM and ASC phases. 
This is more apparent in the $M$ dependence of $\chi_{\rm c} (0)$, as shown in the inset of 
Fig.~\ref{fig:xc_hc}, because for large $M$ 
\begin{equation}
\chi_{\rm c}(0) \propto \exp(\tau_{\rm c}M) \label{eq:chicexp}
\end{equation}
with $\tau_{\rm c}<0$ (also see Table~\ref{tab:exp_hc}), and thus 
\begin{equation}
\lim_{M\to\infty}\chi_{\rm c}(0) \to 0
\end{equation} 
in the LM and ASC phases. 
However, 
for the exact phase boundary at $\varepsilon=\varepsilon_c$, we find that $\chi_{\rm s}(0)$ as 
well as $\chi_{\rm c}(0)$ increases exponentially with $M$ (see the insets of Figs.~\ref{fig:xs_hc} and 
\ref{fig:xc_hc}, and also see Table~\ref{tab:exp_hc}), and therefore 
\begin{equation}
\lim_{M\to\infty}\chi_{\rm s}(0) \to \infty 
\end{equation}
and 
\begin{equation}
\lim_{M\to\infty}\chi_{\rm c}(0) \to \infty, 
\end{equation}
suggesting the VF point. 
In the renormalization group analysis, the VF fixed point is characterized by the renormalized 
parameters ($\varepsilon^{\ast}$, $U^{\ast}$, $V^{\ast}$) = ($0$, $\infty$, $0$) 
for $\varepsilon < U/2$~\cite{gonzalez-buxton}. 
Consequently, three local states at the impurity site, $\left| 0 \right>_i$, $\left| \uparrow \right>_i$, 
and $\left| \downarrow \right>_i$, contribute equally to the threefold-degenerate ground state and thus 
both $\chi_{\rm s}(0)$ and $\chi_{\rm c}(0)$ diverge.

\subsection{\label{sec:pga}Results for the pseudogap Anderson model}

Here, we compare the results of the local static quantities, the ground-state 
phase diagram, and the excitation spectra for the pseudogap Anderson model at half-filling, i.e., $n=1$, 
to those for the single-impurity Anderson model on the honeycomb lattice discussed above. 

\begin{figure}[htbp]
\begin{center}
\includegraphics[width=\hsize]{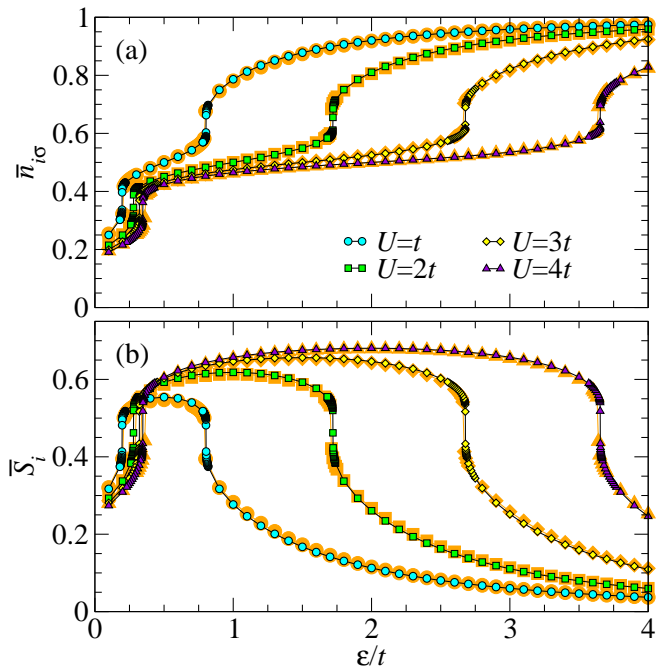}
\caption{(Color online) 
(a) Local density $\bar{n}_{i\sigma}$ and (b) total spin $\bar{S}_{i}$ at the impurity site 
for the pseudogap Anderson model 
as a function of $\varepsilon$ with $V=t$ and various $U$ indicated in (a). 
The logarithmic discretization scheme with $\Lambda = 1.15$ and $M=128$ is used. 
For comparison, the results for the single-impurity Anderson model 
on the honeycomb lattice shown in Fig.~\ref{fig:ns} are also indicated by orange shaded symbols. 
}
\label{fig:ns_pg}
\end{center}
\end{figure}

Figure~\ref{fig:ns_pg} shows the local density $\bar{n}_{i\sigma}$ and total spin $\bar{S}_i$ 
at the impurity site for the pseudogap Anderson model. 
We find in Fig.~\ref{fig:ns_pg} that these local static quantities, including the phase boundaries where 
the abrupt changes of these quantities occur, for the pseudogap Anderson model are 
in quantitatively excellent agreement with those for the single-impurity Anderson model on the honeycomb lattice. 
Since the pseudogap Anderson model is a low-energy effective model for the single-impurity Anderson 
model on the honeycomb lattice, our results ensure that the low-energy effective description 
of the pseudogap Anderson model is quantitatively valid for these static quantities.

\begin{figure}[htbp]
\includegraphics[width=0.9\hsize]{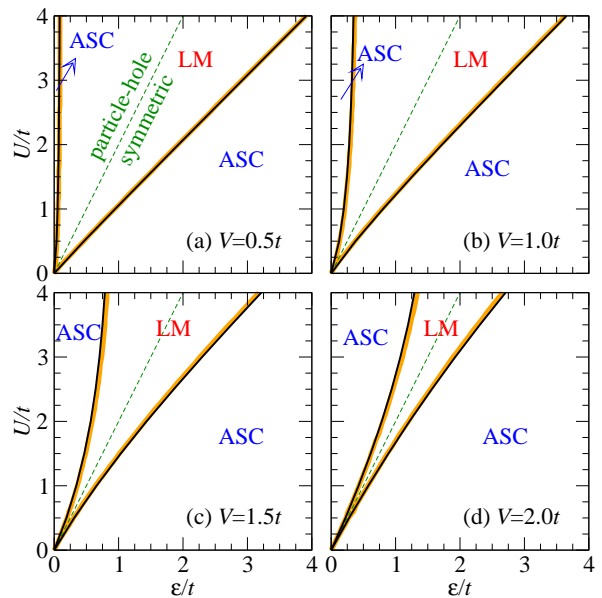}
\caption{(Color online) 
Ground-state phase diagrams for the pseudogap Anderson model 
in a wide range of parameters, 
$\varepsilon$ and $U$, with four different values of $V$ indicated in the figures. 
LM (ASC) stands for the local moment (asymmetric strong coupling) phase. 
The particle-hole symmetric line $\varepsilon  = U/2$ is indicated by green dashed lines. 
The phase boundaries (black solid lines) are determined by the calculations 
of the static quantities shown in Fig.~\ref{fig:ns_pg}.
For comparison, the phase boundaries for the single-impurity Anderson model on the honeycomb lattice 
are also indicated by orange bold lines.
 }
\label{fig:pd_pg}
\end{figure}

Systematically calculating $\bar{n}_{i\sigma}$ and $\bar{S}_i$ for different values of $V$, we 
obtain in Fig.~\ref{fig:pd_pg} the ground-state phase diagram for the pseudogap Anderson model. 
We find that the phase boundaries are almost identical to those for the single-impurity Anderson model 
on the honeycomb lattice. For instance, the VF point for $V=t$ and $U=2t$ is located at 
$\varepsilon_c = 1.721t$--$1.722t$ for the pseudogap Anderson model, 
which is comparable with $\varepsilon_c = 1.707t$--$1.708t$ for the single-impurity Anderson 
model on the honeycomb lattice.

\begin{figure}[tbhp]
\includegraphics[width=\hsize]{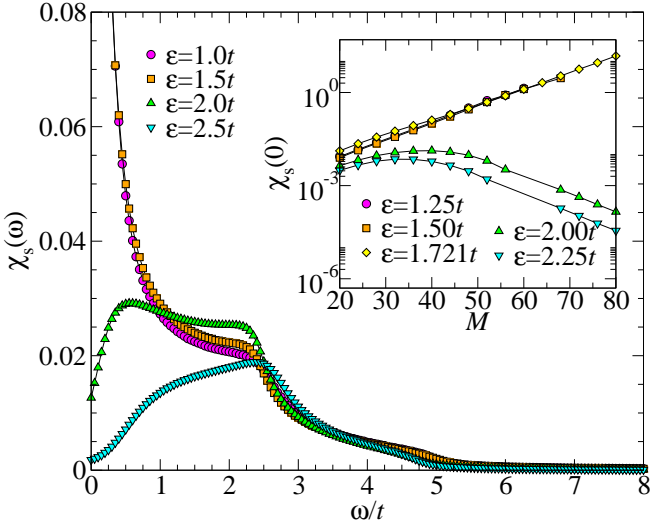}
\caption{(Color online) Spin excitation spectra $\chi_{\rm s}(\omega)$ at the impurity site 
for the pseudogap Anderson model with $V=t$ and $U=2t$. 
The constant discretization scheme with $M=100$ is used and a broadening factor of $\eta = 0.2t$ is set. 
The inset shows the $M$ dependence of $\chi_{\rm s}(0)$ calculated using the logarithmic 
discretization scheme with $\Lambda = 1.15$ and 
$\eta = 25W \Lambda^{-M}$. 
}
\label{fig:xs_pg}
\end{figure}

\begin{figure}[tbhp]
\includegraphics[width=\hsize]{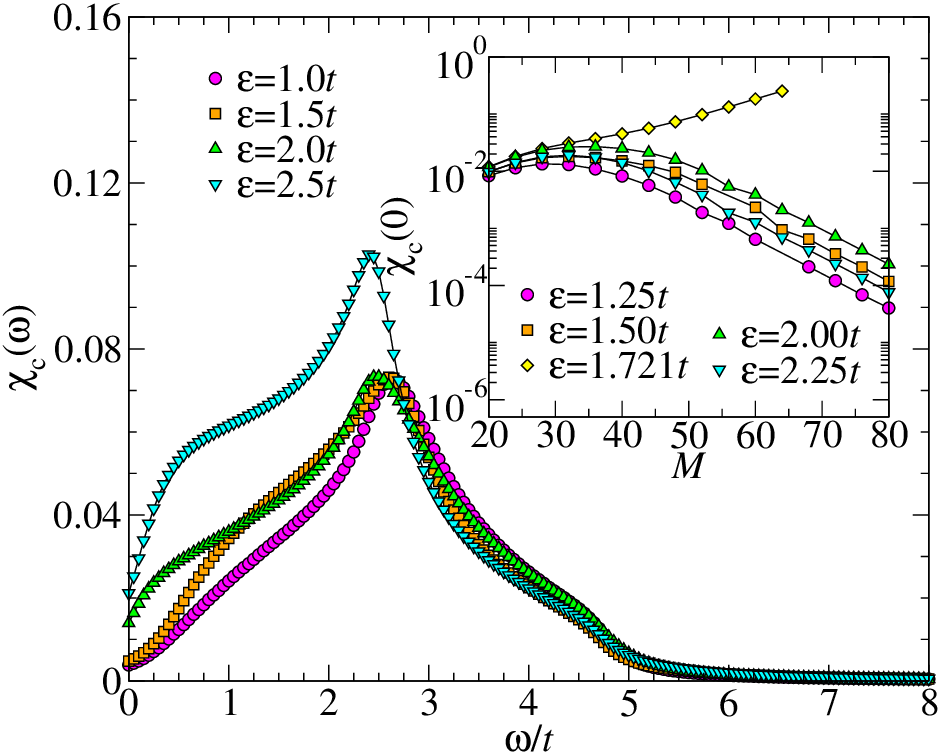}
\caption{(Color online) Charge excitation spectra $\chi_{\rm c}(\omega)$ at the impurity site 
for the pseudogap Anderson model with $V=t$ and $U=2t$. 
The constant discretization scheme with $M=100$ is used and a broadening factor of $\eta = 0.2t$ is set. 
The inset shows the $M$ dependence of $\chi_{\rm c}(0)$ calculated using the logarithmic 
discretization scheme with $\Lambda = 1.15$ and 
$\eta = 25W \Lambda^{-M}$. }
\label{fig:xc_pg}
\end{figure}

Figures~\ref{fig:xs_pg} and \ref{fig:xc_pg} show the spin and charge excitation spectra at the impurity site, 
$\chi_{\rm s}(\omega)$ and $\chi_{\rm c}(\omega)$, respectively. 
Comparing with Figs.~\ref{fig:xs_hc} and \ref{fig:xc_hc}, 
the line shapes of these excitation spectra are apparently different from those for the 
single-impurity Anderson model on the honeycomb lattice. 
However, as shown in the insets of Figs.~\ref{fig:xs_pg} and \ref{fig:xc_pg}, we find that the asymptotic 
behavior of these quantities around $\omega\sim0$ are qualitatively the same for both models, 
i.e., $\chi_{\rm s} (0) \to \infty$ and $\chi_{\rm c} (0) \to 0$ in the LM phase,  
$\chi_{\rm s} (0) \to 0$ and $\chi_{\rm c} (0) \to 0$ in the ASC phase, 
and $\chi_{\rm s} (0) \to \infty$ and $\chi_{\rm c} (0) \to \infty$ at the VF point. 
Assuming that $\chi_{\rm s} (0)$ and $\chi_{\rm c} (0)$ diverge or decay exponentially for large $M$, as in 
Eqs.~(\ref{eq:chisexp}) and (\ref{eq:chicexp}), 
we can estimate the diverging or decaying factors $\tau_{\rm s}$ and $\tau_{\rm c}$ 
for the pseudogap Anderson model. 
As shown in Table~\ref{tab:exp_pg}, we find that the obtained $\tau_{\rm s}$ and $\tau_{\rm c}$ are indeed 
very close to those for the single-impurity Anderson model on the honeycomb lattice shown in Table~\ref{tab:exp_hc}, 
indicating that the asymptotically low-energy excitations of the single-impurity Anderson model on the 
honeycomb lattice are also well described by the low-energy effective pseudogap Anderson model.

\begin{table}[htbp]
\caption{
The diverging or decaying factors $\tau_{\rm s}$ and $\tau_{\rm c}$ of the local spin 
and charge excitation spectra at $\omega=0$, $\chi_{\rm s}(0)\propto\exp(\tau_{\rm s}M)$ 
and $\chi_{\rm c}(0)\propto\exp(\tau_{\rm c}M)$, respectively, 
for large $M$ in the three different regions of the phase diagram, i.e, the LM and ASC phases and the VF point, 
estimated from the results shown in Figs.~\ref{fig:xs_pg} and \ref{fig:xc_pg}. }
\label{tab:exp_pg}
\begin{tabular}{c|ccc}
\hline \hline
   {}   &  LM phase & ASC phase & VF point \\
      \hline
$\tau_{\rm s}$  & $0.132$ & $-0.140$ & $0.115$ \\
$\tau_{\rm c}$  & $-0.140$ & $-0.140$ & $0.079$ \\
\hline \hline
\end{tabular}
\end{table}

\subsection{\label{entanglement}Entanglement spectrum} 

Let us now discuss the entanglement spectrum for the ground state of the single impurity Anderson model 
on the honeycomb lattice. 
In the DMRG method, the system is divided into two parts, the left and right blocks, as shown in 
Fig.~\ref{fig:model}(b), 
with the sizes being $l_{\rm L}$ and $l_{\rm R}$, respectively, i.e., $L=l_{\rm L}+l_{\rm R}$. 
Thus, the wave function is generally represented as 
\begin{eqnarray}
\left| \psi \right> = \sum_{i}\sum_{j} \psi_{i,j} \left| i \right>_{\rm L} \otimes \left| j \right>_{\rm R}, 
\end{eqnarray}
where $\left| i \right>_{\rm L}$ and $\left| j \right>_{\rm R}$ indicate the bases of the left and right blocks, respectively. 
The reduced density matrix $\hat{\rho}_{\rm L}$ for the left block is 
\begin{eqnarray}
(\hat{\rho}_{\rm L})_{i,i^{\prime}} = \sum_{j} \psi_{i,j} \psi_{i^{\prime},j}^{\ast}, 
\end{eqnarray}
and the $k$th eigenvalue of $\hat\rho_{\rm L}$ is denoted as $\lambda_k$ 
in descending order, i.e., 
\begin{eqnarray}
\lambda_1 \geq \lambda_2  \geq \cdots\geq\lambda_{m_{\rm D}}, 
\end{eqnarray}
where $m_{\rm D}$ is the number of density matrix eigenstates kept in the DMRG calculations. 
One can readily show that $0\le\lambda_k\le1$ and $\sum_k\lambda_k=1$ 
when $\langle\psi|\psi\rangle=1$.
The entanglement spectrum $\xi_k$~\cite{li} is defined as
\begin{eqnarray}
\xi_k = -\ln \lambda_k, 
\end{eqnarray}
and hence
\begin{eqnarray}
\xi_1 \leq \xi_2 \leq  \cdots \leq \xi_{m_{\rm D}}. 
\end{eqnarray}

\begin{figure}[htbp]
\includegraphics[width=\hsize]{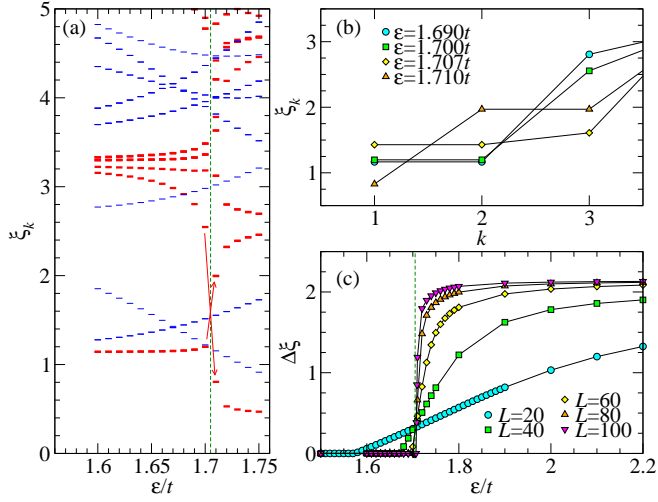}
\caption{(Color online) 
Entanglement spectrum for the ground state of the single-impurity Anderson model 
on the honeycomb lattice 
with $V=t$ and $U=2t$. The logarithmic discretization scheme with $\Lambda = 1.15$ is used.
(a) Low-lying entanglement spectrum $\xi_k$ (red bars) for $\varepsilon$ 
at the vicinity of the phase boundary $\varepsilon_c\sim1.707t$--$1.708t$ (dashed vertical line) 
determined in Fig.~\ref{fig:ns}. 
The calculations are for $L=100$ with the left block size $l_{\rm L} = 51$. 
For comparison, the results for $L=40$ with $l_{\rm L} = 21$ are also shown by 
blue bars, in which the phase boundary is approximately $1.67t$. 
Red arrows are guide for eyes. 
(b) The lowest three levels of $\xi_k$ for $\varepsilon$ close to $\varepsilon_c$, calculated 
for $L=100$ with $l_{\rm L} = 51$. 
(c) Gap of the entanglement spectrum, $\Delta \xi = \xi_2 - \xi_1$, as a function of $\varepsilon$ for various 
$L$ with $l_{\rm L} = L/2 + 1$. 
The phase boundary $\varepsilon_c$ determined in Fig.~\ref{fig:ns} is indicated 
by a dashed vertical line.
}
\label{fig:espec}
\end{figure}

Figure~\ref{fig:espec} shows a low-lying part of $\xi_k$ for the ground state
at the vicinity of the phase boundary. 
It is clear in Fig.~\ref{fig:espec}(a) that the lowest and the first excited entanglement levels cross 
at $\varepsilon = \varepsilon_c$. 
As shown in Fig.~\ref{fig:espec}(b), we find 
that the lowest entanglement level for $l_{\rm L} $ odd is doubly degenerate in the LM phase, 
singlet in the ASC phase, and accidentally three fold degenerate at 
$\varepsilon=\varepsilon_c$, i.e., the VF point~\cite{note7}. 
Because of the qualitatively different behavior, we can consider the gap of the entanglement 
spectrum 
\begin{equation}
\Delta \xi = \xi_2 - \xi_1
\end{equation} 
as an ``order parameter" to distinguish the different phases in the phase diagram. 
As shown in Fig.~\ref{fig:espec}(c), $\Delta \xi$ changes abruptly at $\varepsilon=\varepsilon_c$ 
for large $M$ and it is finite only when $\varepsilon>\varepsilon_c$. 
Indeed, the phase boundary determined from $\xi_k$ is the same as the one estimated 
in Fig.~\ref{fig:ns}. 
This clearly demonstrates that $\Delta \xi$ serves as 
a quantity to determine the phase boundary of the impurity quantum phase transition. 
We should emphasize here that the ground state is always singlet as long as 
$L$ is even and finite. Therefore, the similar characteristic feature of the degeneracy 
in the low-lying entanglement spectrum is absent in the low-lying energy spectrum.

\begin{figure}[thbp]
\begin{center}
\includegraphics[width=\hsize]{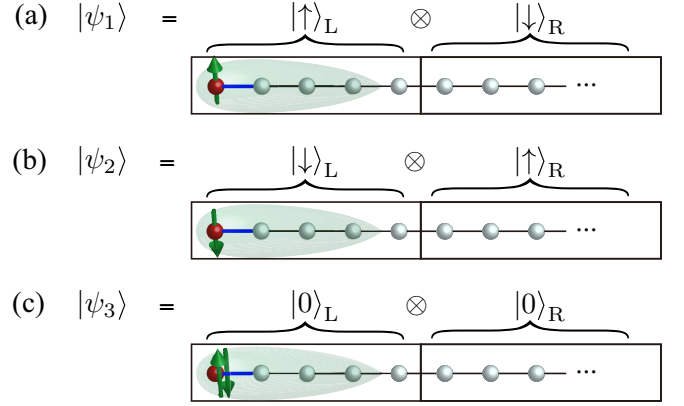}
\caption{(Color online) 
(a) One of the doubly degenerate ground states for $L\to\infty$, $|\psi_1\rangle$, in the LM phase. 
(b) The other state, $\left| \psi_2 \right>$, of the doubly degenerate ground states in the LM phase. 
(c) The singlet ground state $\left| \psi_3 \right>$ in the ASC phase. 
Here, $s=\uparrow$, $\downarrow$, and 0 in $|s\rangle_{\rm L(R)}$ represents the $z$ component of  
total spin $S_z=1/2$, $-1/2$, and 0, respectively, in the left (right) block of 
$|\psi_i\rangle$ for $i=1$, 2, and 3. 
Red spheres at the left edge represent the impurity site and green arrows indicate 
the local spin configurations around the impurity site which may be spatially extended 
into the green shaded region. 
}
\label{fig:st}
\end{center}
\end{figure}

Let us now discuss the intuitive understanding of the origin for the different degeneracy of 
the low-lying entanglement spectrum $\xi_k$ in each phase. 
We first note that in our calculations the impurity site is located at the left edge  
[see Fig.~\ref{fig:model}(b)] and the degeneracy of the lowest entanglement level in the 
LM phase occurs only 
for $l_{\rm L}$ odd. This implies that the degeneracy in the LM phase is due to 
the quantum number conservation in each block. 
Clearly, the ground state of the LM phase is doubly degenerate in $L\to\infty$ and 
is described schematically as 
$\left| \psi_1 \right> \sim \left| \uparrow \right>_{\rm L} \otimes \left| \downarrow \right>_{\rm R}$ 
and $\left| \psi_2 \right> \sim \left| \downarrow \right>_{\rm L} \otimes \left| \uparrow \right>_{\rm R}$, 
where 
$s=\uparrow$, $\downarrow$, and 0 in $|s\rangle_{\rm L(R)}$ indicates the $z$ component of  
total spin $S_z=1/2$, $-1/2$, and 0, respectively, 
in the left (right) block of the ground state. 
Here, nonzero $s $ in $ \left| s \right>_{\rm L}$ is due to the localized spin formed around the impurity site, 
and correspondingly $ \left| s \right>_{\rm R}$ has the opposite spin to compensate the spin in the left block 
[see Figs.~\ref{fig:st}(a) and \ref{fig:st}(b)]. 
In a finite $L$, however, these two states $\left| \psi_1 \right>$ and $\left| \psi_2 \right>$ 
are entangled and the ground state is 
$ \left| \psi_0 \right> \sim (\left| \psi_1 \right> - \left| \psi_2 \right>)/\sqrt{2}$.
We can now readily show that the lowest $\xi_k$ is doubly degenerate.
This is no longer the case when $l_{\rm L}$ is even. Although the impurity site is still represented approximately as 
$\left|\uparrow\right>_i$ or $\left|\downarrow\right>_i$, $\left|s\right>_{\rm L}\sim \left|0\right>_{\rm L}$ for 
$l_{\rm L}$ even 
because $\left| \uparrow \right>_i$ and $\left| \downarrow \right>_i$ must be entangled 
in the left block due to the conservation of $S_z$~\cite{note9}.   
Therefore, the lowest $\xi_k$ is no longer degenerate when $l_{\rm L}$ is even.

In the ASC phase for $\varepsilon > U/2$, the impurity site is approximately 
doubly occupied $\left| \uparrow \downarrow \right>_i$. Therefore, the ground state is described as  
$\left| \psi_3 \right> \sim \left| 0 \right>_{\rm L} \otimes \left| 0 \right>_{\rm R}$, 
as schematically shown in Fig.~\ref{fig:st}(c), 
and thus the lowest $\xi_k$ is not degenerate~\cite{note8}. 
The VF point, on the other hand, corresponds to the special case where 
$|\psi_1\rangle$, $|\psi_2\rangle$, and $|\psi_3\rangle$ are all degenerate in $L\to\infty$. 
These three states are entangled in a finite $L$ and 
the ground state is represented approximately as 
$\left| \psi_0 \right> \sim (\left| \psi_1 \right> - \left| \psi_2 \right> + \left| \psi_3 \right> )/\sqrt{3}$.
We can now show that  the lowest $\xi_k$ for $|\psi_0\rangle$ is three fold degenerate.

\begin{figure}[thbp]
\begin{center}
\includegraphics[width=\hsize]{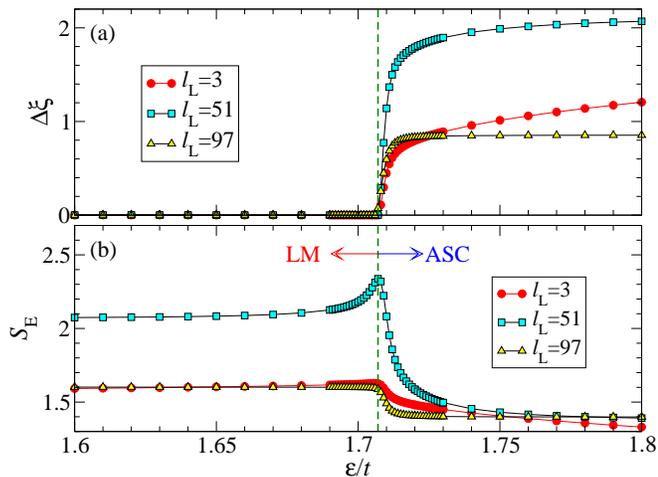}
\caption{(Color online) 
(a) Entanglement spectrum gap $\Delta \xi$ and (b) entanglement entropy $S_{\rm E}$ 
for the ground state of the single-impurity Anderson model 
on the honeycomb lattice with $V=t$ and $U=2t$. 
The calculations are for $L=100$ with three different $l_{\rm L}$ indicated in the figures. 
The logarithmic discretization scheme with $\Lambda = 1.15$ and $M=100$ is used.
Green dashed lines indicate the phase boundary $\varepsilon_c$ determined in 
Fig.~\ref{fig:ns}. 
}
\label{fig:ee}
\end{center}
\end{figure}

Finally, let us briefly discuss the $l_{\rm L}$ dependence of the 
entanglement spectrum gap $\Delta \xi$ and the entanglement entropy $S_{\rm E}$. 
The entanglement entropy is a quantity to measure the degree 
of quantum entanglement between the left and right blocks of a given quantum state $|\psi\rangle$ 
and is defined as 
\begin{eqnarray}
S_{\rm E} = - {\rm Tr} \hat{\rho}_{\rm L} \ln \hat{\rho}_{\rm L} = - \sum_{k=1}^{m_{\rm D}} \lambda_k \ln \lambda_k. 
\end{eqnarray}
Fig.~\ref{fig:ee} shows the $l_{\rm L}$ dependence of these quantities calculated for 
the ground state of the single-impurity Anderson model on the honeycomb lattice.

Because the degeneracy of the lowest entanglement level, i.e., the largest $\lambda_k$, is 
different in each region of the phase diagram, we find in Fig.~\ref{fig:ee}(b) that $S_{\rm E}$ can 
exhibit a maximum around the phase boundary when $l_{\rm L} $ is chosen appropriately. 
Therefore, $S_{\rm E}$ can also be an indicator to estimate the transition point of the impurity 
quantum phase transition. However, 
in contrast to $\Delta\xi$, the variation of $S_{\rm E}$ is rather smooth across the transition 
for a finite $L$. 
Furthermore, we find that $S_{\rm E}$ can even monotonically decrease with increasing 
$\varepsilon$ without showing a peak structure around the transition point 
[see, for example, the results for $l_{\rm L}=97$ in Fig.~\ref{fig:ee}(b)]. 
This indicates that the maximum of $S_{\rm E}$ is not always located at the phase boundary. 
Thus, the degeneracy of the low-lying entanglement spectrum is a much better quantity 
to determine the phase boundary for the finite-$L$ calculations.

\section{\label{summary}Summary}

We have introduced the DMRG method in energy space for Anderson impurity models, 
which allows us for calculations in the thermodynamic limit. We have   
applied this method to the single-impurity Anderson model on the honeycomb lattice 
to establish the ground-state phase diagram at half-filling.  
By systematically calculating the local static quantities, we have found that the phase 
diagram contains two phases, i.e., the LM phase and the ASC phase, 
but no Kondo screening phase.  
To support these results, we have also calculated the spin and charge excitation spectra 
at the impurity site, which behave qualitatively differently in these phases and reveal the 
existence of the VF point at the phase boundary. These results are thus qualitatively 
in good agreement with those obtained previously by the low-energy approximate approaches. 

For quantitative comparison, we have also studied the low-energy effective pseudogap Anderson model 
using the method introduced here. 
Although the high-energy excitations are obviously different, we have found that the ground-state 
phase diagram and the asymptotic low-energy excitations are in good quantitative agreement with those for 
the single-impurity Anderson model on the honeycomb lattice. 
Therefore, our result provides the first quantitative justification for studies based on 
the low-energy effective models.

We have also discussed the entanglement properties for the ground state of the 
single-impurity Anderson model on the honeycomb lattice. We have found that the low-lying 
entanglement spectrum exhibits qualitatively different behaviors in the different regions of the 
phase diagram: the lowest entanglement 
level is doubly degenerate for the LM phase, singlet for the ASC phase, and three fold degenerate 
at the VF point. We have also provided the intuitive understanding of these different behaviors in the 
degeneracy of the lowest entanglement level. 
The degeneracy of the lowest entanglement level differs 
from the degeneracy of the lowest-energy level because the ground state is found 
to be always singlet as long as $L$ is even and finite. 
Furthermore, we have shown that the entanglement entropy can exhibit a broad 
maximum around the phase transition point when the ground state is properly separated 
to calculate the entanglement entropy. However, this is not always the case and sometimes the 
entanglement entropy varies monotonically across the transition. Therefore, we conclude that 
the entanglement spectrum is a better quantity to distinguish different phases in the impurity 
quantum phase transition.

Finally, our present analysis has no intention to make the quantitative comparison 
with experiments on the impurity problem in graphene~\cite{maccreary,nair}. 
For the quantitative comparison, further details not included in the simplest single-impurity Anderson 
model should be considered. 
For example, the electron correlation in the conduction band might have a significant effect on the nature 
of quasiparticles~\cite{Meng2010,Sorella2012,Assaad2013,Toldin2015,Otsuka2015}. 
The incorporation of different chemical bondings between 
graphene and adatom inevitably requires more complex hybridization, which  
affects the local electronic and magnetic properties around the impurity~\cite{han}. 
Furthermore, the spin-orbit coupling induced by the structural deformation around the impurity can be significantly 
large~\cite{neto2, schmidt, weeks, zhou, balakrishnan}. A transition metal substrate can also 
induce a giant Rashba splitting in graphene~\cite{marchenko, mastrogiuseppe}. 
The method introduced here, 
in combination with the first-principles band-structure calculation 
based on the density functional theory, 
would be a valuable extension to disentangle these effects 
for the impurity problem in graphene.

\section*{Acknowledgments}
The authors are grateful to K. Shinjo, R. Peters, E. Minamitani, and H. Watanabe for valuable discussions. 
The computation has been done using the RIKEN Cluster of Clusters (RICC), 
the RIKEN supercomputer system (HOKUSAI GreatWave), and the facilities at Supercomputer Center 
in ISSP, Information Technology Center, University of Tokyo. 
This work has been supported by Grants-in-Aid for Scientific Research from JSPS under 
the Grants No. 24740269 and No. 26800171 and in part by RIKEN iTHES Project and Molecular Systems.

\appendix

\section{\label{app:hr}
Effective actions of the impurity site for $\mathcal{H}_{\rm AIM}$, 
$\mathcal{H}_{\omega}$, and $\mathcal{H}_r$. }
In this appendix, we show that the effective action of the impurity site 
for the single-impurity Anderson model $\mathcal{H}_{\rm AIM}$ on the honeycomb lattice 
is exactly the same as those for $\mathcal{H}_{\omega}$ and $\mathcal{H}_r$ in energy space.

In the momentum space, $\mathcal{H}_c$ and $\mathcal{H}_V$ in $\mathcal{H}_{\rm AIM}$ are written, 
respectively, as 
\begin{eqnarray}
\mathcal{H}_c = \sum_{\bf k} \sum_{\sigma = \uparrow, \downarrow} \sum_{\zeta = \pm} \varepsilon_{\bf k}^{(\zeta)} 
c_{{\bf k},\zeta,\sigma}^{\dagger} c_{{\bf k},\zeta,\sigma} 
\end{eqnarray}
and 
\begin{eqnarray}
\mathcal{H}_V = \frac{V}{\sqrt{2N}} \sum_{\bf k} \sum_{\sigma=\uparrow,\downarrow} \sum_{\zeta = \pm}
e^{{\rm i}{\bf k}\cdot {\bf r}_0} c_{i,\sigma}^{\dagger} c_{{\bf k},\zeta,\sigma} + {\rm H.c.}, 
\end{eqnarray}
where $N$ is the number of unit cells and $c_{{\bf k},\zeta,\sigma}^\dag$ ($c_{{\bf k},\zeta,\sigma}$) 
is an electron creation (annihilation) operator of the conduction band at momentum ${\bf k}$ with the band dispersion 
\begin{eqnarray}
\varepsilon_{\bf k}^{(\zeta)} = \zeta t \left| 1 + e^{{\rm i}{\bf k}\cdot {\bf a}_1} + e^{{\rm i}{\bf k} \cdot {\bf a}_2} \right|. 
\end{eqnarray}
Here, ${\bf a}_1$ and ${\bf a}_2$ are the primitive lattice vectors of the honeycomb lattice. 
The partition function for $\mathcal{H}_{\rm AIM}$ is then 
\begin{eqnarray}
Z_{\rm AIM} &=& \int \mathcal{D}\bar{\psi} \mathcal{D} \psi \exp \left[ - S_{\rm AIM} \right], 
\end{eqnarray}
where 
\begin{eqnarray}
S_{\rm AIM} &=& \int_0^{\beta} {\rm d}\tau \sum_{\sigma = \uparrow,\downarrow} 
\bar{\psi}_{i,\sigma}(\tau) (\partial_{\tau} - \varepsilon ) \psi_{i,\sigma}(\tau) \nonumber \\
& +& \int_0^{\beta} {\rm d}\tau \sum_{{\bf k}} \sum_{\sigma = \uparrow,\downarrow} \sum_{\zeta = \pm}
\bar{\psi}_{{\bf k},\zeta,\sigma}(\tau) \partial_{\tau} \psi_{{\bf k},\zeta,\sigma}(\tau) \nonumber \\
& + &\int_0^{\beta} {\rm d}\tau \sum_{\sigma = \uparrow,\downarrow} \sum_{\bf k} \sum_{\zeta = \pm} 
\varepsilon_{\bf k}^{(\zeta)} \bar{\psi}_{{\bf k},\zeta,\sigma}(\tau) \psi_{{\bf k},\zeta,\sigma}(\tau) \nonumber \\
& +& \frac{V}{\sqrt{2N}} \int {\rm d}\tau \sum_{\sigma = \uparrow,\downarrow} \sum_{\bf k} 
\left [ e^{{\rm i}{\bf k} \cdot {\bf r}_0}
 \bar{\psi}_{i,\sigma}(\tau) \psi_{{\bf k},\zeta,\sigma}(\tau) \right. \nonumber \\
 &&\quad\quad\quad\quad\quad\quad\quad\quad\quad 
\left. +    e^{-{\rm i}{\bf k} \cdot {\bf r}_0}   \bar{\psi}_{{\bf k},\zeta,\sigma}(\tau) \psi_{i,\sigma}(\tau) \right]  
\nonumber \\
& + &U \int_0^{\beta} {\rm d} \tau \bar{\psi}_{i,\uparrow}(\tau) \bar{\psi}_{i,\downarrow}(\tau) \psi_{i,\downarrow}(\tau) \psi_{i,\uparrow}(\tau). 
\end{eqnarray}
Here, $\psi_{i,\sigma} ( \tau )$ [$ \bar{\psi}_{i,\sigma} (\tau )$] 
and $ \psi_{{\bf k},\zeta,\sigma}(\tau)$ [$ \bar{\psi}_{{\bf k},\zeta,\sigma}(\tau)$] 
are the Grassman's numbers corresponding to $c_{i,\sigma}$ ($c_{i,\sigma}^{\dagger}$) and 
$c_{{\bf k},\zeta,\sigma}$ ($c_{{\bf k},\zeta,\sigma}^{\dagger}$), respectively, at imaginary time $\tau$, and 
$\beta$ is the inverse temperature. 
Carrying out the Gaussian integrals over the Grassman's numbers for the conduction electrons, 
we obtain 
\begin{eqnarray}
Z_{\rm AIM} &=& C \int \mathcal{D} \bar{\psi}_i \mathcal{D} \psi_i \exp \left[ - S_{\rm imp} \right], 
\end{eqnarray}
where the effective action $S_{\rm imp}$ of the impurity site is given as 
\begin{eqnarray}\label{eq:simp}
S_{\rm imp} &=&  \frac{1}{\beta} \sum_{n=-\infty}^{\infty} \sum_{\sigma = \uparrow,\downarrow} 
\bar{\psi}_{n,i,\sigma} [ -{\rm i}\omega_n - \varepsilon + \Delta ( {\rm i} \omega_n )] \psi_{n,i,\sigma} \nonumber \\
& +& U \int_0^{\beta} {\rm d} \tau \bar{\psi}_{i,\uparrow}(\tau) \bar{\psi}_{i,\downarrow}(\tau) \psi_{i,\downarrow}(\tau) \psi_{i,\uparrow}(\tau) 
\end{eqnarray}
with $\omega_n=(2n+1)\pi/\beta$ ($n$: integer) and $C$ being a constant. 
Here, we have introduced that 
\begin{eqnarray}
\psi_{n,i,\sigma} &=& \int_0^{\beta} {\rm d}\tau e^{{\rm i}\omega_n \tau } \psi_{i,\sigma} (\tau), \\
\bar{\psi}_{n,i,\sigma} &=& \int_0^{\beta} {\rm d}\tau e^{-{\rm i}\omega_n \tau} \bar{\psi}_{i,\sigma} (\tau), 
\end{eqnarray}
and 
\begin{eqnarray}
\Delta ( {\rm i} \omega_n ) &=& \frac{V^2}{2N} 
\sum_{\bf k} \sum_{\zeta = \pm} \frac{1}{{\rm i}\omega_n - \varepsilon_{\bf k}^{(\zeta)} } \nonumber \\
&=& V^2 \int_{-\infty}^{\infty} d\omega \frac{\rho ( \omega )}{{\rm i}\omega_n - \omega }, \label{eq:hyb}
\end{eqnarray}
where $\rho ( \omega )$ is the local density of states per spin for the conduction band.

Similarly, the partition function for $\mathcal{H}_{\omega}$ in Eq.~(\ref{eq:hw}) is given as
\begin{eqnarray}
Z_{\omega} &=& \int \mathcal{D}\bar{\psi} \mathcal{D} \psi \exp \left[ - S_{\omega} \right], 
\end{eqnarray}
where 
\begin{eqnarray}
S_{\omega} &=& \int_0^{\beta} {\rm d} \tau \sum_{\sigma = \uparrow,\downarrow} 
\left[ \bar{\psi}_{i,\sigma}(\tau) ( \partial_{\tau} - \varepsilon ) \psi_{i,\sigma}(\tau) \right. \nonumber \\
&& \quad\quad\quad\quad\quad\quad  \left. + \int {\rm d}\omega
\bar{\psi}_{\omega,\sigma}(\tau) \partial_{\tau} \psi_{\omega,\sigma}(\tau) \right] \nonumber \\
& +& \int_0^{\beta} d\tau \sum_{\sigma=\uparrow,\downarrow} \int {\rm d}\omega\, \omega\, 
\bar{\psi}_{\omega,\sigma}(\tau) \psi_{\omega,\sigma}(\tau) \nonumber \\
& +& V \int d\tau \sum_{\sigma = \uparrow,\downarrow} \int {\rm d} \omega \sqrt{\rho(\omega )} 
\left[ \bar{\psi}_{i,\sigma}(\tau) \psi_{\omega,\sigma}(\tau) \right. \nonumber \\
&&\quad\quad\quad\quad\quad\quad\quad\quad\quad\quad\quad\quad  \left. + \bar{\psi}_{\omega, \sigma}(\tau) \psi_{i,\sigma}(\tau) \right] \nonumber \\
& +& U \int_0^{\beta} {\rm d} \tau \bar{\psi}_{i,\uparrow}(\tau) \bar{\psi}_{i,\downarrow}(\tau) \psi_{i,\downarrow}(\tau) \psi_{i,\uparrow}(\tau) 
\end{eqnarray}
and $ \psi_{\omega,\sigma}(\tau)$ [$ \bar{\psi}_{\omega,\sigma} ( \tau )$] 
is the Grassmann's number corresponding to $a_{\omega,\sigma}$ ($a_{\omega,\sigma}^{\dagger}$) at imaginary 
time $\tau$. 
Carrying out the Gaussian integral for the conduction band, we can readily show that 
the effective action $S_{\rm imp}$ of the impurity site is exactly the same as the one for $\mathcal H_{\rm AIM}$ 
with the same $\Delta ( {\rm i} \omega_n )$ given in Eq.~(\ref{eq:hyb}). 
Therefore, as long as the impurity properties are considered, these two models described by Hamiltonians 
$\mathcal{H}_{\rm AIM}$ and $\mathcal{H}_{\omega}$ are equivalent. 

We can follow the same analysis to obtain the effective action of the impurity site for $\mathcal{H}_r$ 
given in Eq.~(\ref{eq:hr}), and find that the effective action is exactly the 
same as the one in Eq.~(\ref{eq:simp}) except that $\Delta ( {\rm i} \omega_n ) $ is now replaced by 
\begin{eqnarray}
\Delta_r ( {\rm i} \omega_n ) &=& \sum_{m=1}^M \frac{(\gamma_m^+)^2}{{\rm i}\omega_n - \xi_m^+} + \sum_{m=1}^M 
\frac{(\gamma_m^-)^2}{{\rm i}\omega_n - \xi_m^-} \\
&=& \sum_{m=1}^M \frac{V^2 }{{\rm i}\omega_n - \xi_m^+} \int_{\omega_{m}}^{\omega_{m-1}}{\rm d} \omega \rho(\omega) \nonumber \\
&{}& +
\sum_{m=1}^M \frac{V^2 }{{\rm i}\omega_n - \xi_m^-}  \int_{-\omega_{m-1}}^{-\omega_{m}}{\rm d}\omega \rho(\omega). \label{eq:hyb_dis}
\end{eqnarray}
Comparing Eqs.~(\ref{eq:hyb}) and (\ref{eq:hyb_dis}), 
we can find that 
\begin{equation}
\lim_{M\to\infty} \Delta_r({\rm i}\omega_n) = \Delta({\rm i}\omega_n),
\end{equation}
provided that $\Lambda\to1^+$ is also taken for the logarithmic discretization scheme.
Therefore, the effective action $S_{\rm imp}$ of the impurity site for $\mathcal{H}_r$ becomes exactly 
the same as the one for ${\cal H}_{\rm AIM}$ when the small enough energy interval is adopted 
with $M\to\infty$ and also $\Lambda\to1^+$ for the logarithmic discretization scheme.

\section{\label{app:nd}Further technical details }

In this appendix, we examine further technical details. First, we discuss the $l$ dependence of 
hopping $t_l$ in $\mathcal H$ [Eq.~(\ref{eq:hamchain})] for the single-impurity Anderson model 
on the honeycomb lattice. 
Next, we analyze the convergence behavior of static quantities with respect to $L$ and $\Lambda$. 
We also determine the phase boundary in the thermodynamic limit 
by explicitly taking the limits of $L\to \infty$ and $\Lambda\to 1^+$, which turns out to coincide 
within our numerical accuracy with the phase boundary obtained by the calculations for 
$L=128$ and $\Lambda=1.15$ in Fig.~\ref{fig:ns}. 
Finally, we discuss the convergence issue of dynamical quantities.

\subsection{The $l$ dependence of hopping $t_l$ in $\mathcal H$}

\begin{figure}[htbp]
\includegraphics[width=\hsize]{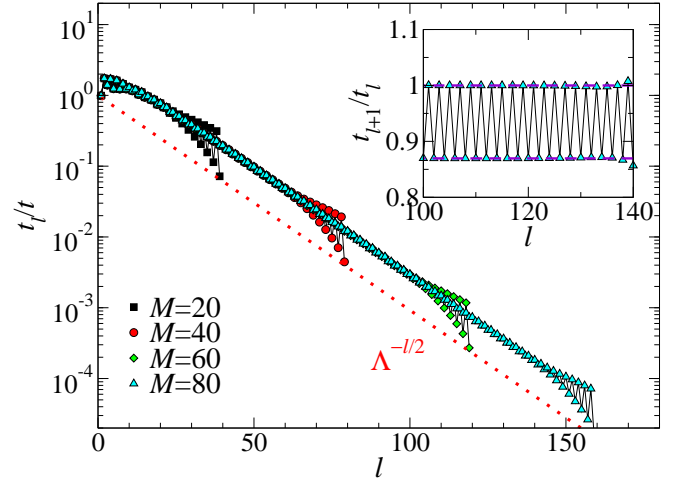}
\caption{(Color online)  $l$ dependence of $t_l$ in $\mathcal H$ for the single-impurity Anderson model 
on the honeycomb lattice with various values of $M$ indicated in the figure. 
Note that the maximum value of $l$ for a given $M$ is $2M+1$.
The logarithmic discretization scheme with $\Lambda = 1.15$ is used.  
A red dotted line indicates $\Lambda^{-l/2}$ with $\Lambda = 1.15$. 
The inset shows $t_{l+1}/t_l$ for $M=80$. 
Two horizontal dashed lines in the inset are $t_{l+1}/t_l=1$ and $\Lambda^{-1}$. 
}
\label{fig:tmat_log}
\end{figure}

Let us first show in Fig.~\ref{fig:tmat_log} the $l$ dependence of $t_l$, 
i.e., the nearest-neighbor hopping between the conduction sites in energy space described by $\mathcal H$, 
for the single-impurity Anderson model 
on the honeycomb lattice when the logarithmic discretization scheme is used. 
It is observed in Fig.~\ref{fig:tmat_log} that $t_l$ decays exponentially with increasing $l$. Indeed, 
we find that $t_l$ decays approximately as $t_l \approx \Lambda^{-l/2}$ except for the oscillatory behavior towards 
one of the edges of the chain opposite to the impurity site [see also Fig.~\ref{fig:model}(b)]. 
Although this oscillatory behavior of $t_l$ is not a major problem for our calculations, we terminate $l$ at 
$L=M$ to save the computational time. As discussed below, this is rationalized because 
the physical quantities are already well converged with $L=M$ as long as $M$ is large enough.

The overall behavior of $t_l$ found in Fig.~\ref{fig:tmat_log} is in good agreement 
with the one for the pseudogap Anderson model. It is known that the asymptotic 
behavior of $t_l$ for the pseudogap Anderson model is 
\begin{eqnarray}
t_l \sim \left\{ 
\begin{array}{ll}
C (\Lambda) \Lambda^{-l/2} & (l:\ {\rm even}), \\
C (\Lambda) \Lambda^{-(l+1)/2} & (l:\ {\rm odd}), \\
\end{array}\right. ,
\end{eqnarray}
where $C(\Lambda)$ is a constant depending only on $\Lambda$~\cite{bulla1}. 
Therefore, $t_{l+1}/t_l \sim \Lambda^{-1}$ for $l$ even and $t_{l+1}/t_l \sim 1$ for $l$ odd. 
As shown in the inset of Fig.~\ref{fig:tmat_log}, we find that $t_l$ for the single-impurity Anderson model 
on the honeycomb lattice also shows the same asymptotic behavior.

\begin{figure}[htbp]
\includegraphics[width=\hsize]{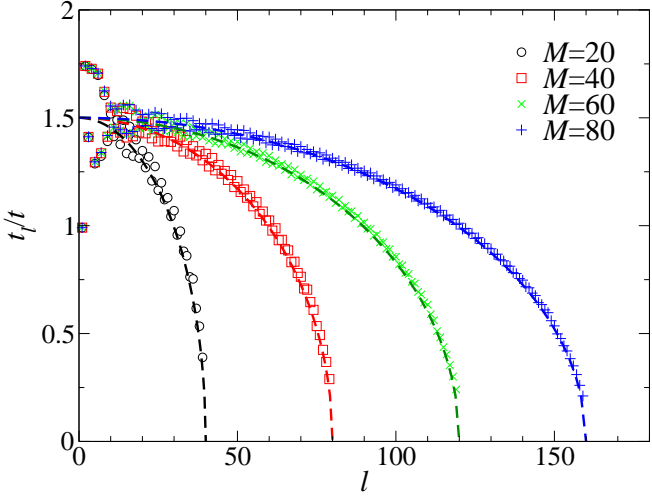}
\caption{(Color online) 
Same as in Fig.~\ref{fig:tmat_log} but 
the constant discretization scheme is used. 
Dashed lines along the symbols indicate $W\sqrt{1-(l/M)^2}/4$. 
}
\label{fig:tmat_const}
\end{figure}

In contrast, the $l$ dependence of $t_l$ is qualitatively different when the constant discretization scheme is used. 
As shown in Fig.~\ref{fig:tmat_const}, $t_l$ is significantly different when different $M$ is used. 
This is simply because 
$\omega_{m}^{\pm}$ for a given $m$ depends directly on $M$ for the constant discretization 
scheme, while it is independent of $M$ for the logarithmic discretization scheme. 
We also find in Fig.~\ref{fig:tmat_const} that the overall behavior of $t_l$ is well described as 
$t_l \approx W\sqrt{1-(l/M)^2}/4$. This is a universal feature when the constant discretization scheme is used since 
the same overall behavior is found even when the constant density of states is assumed.

\subsection{The convergence behavior of static quantities with respect to $L$ and $\Lambda$}

\begin{figure}[htbp]
\includegraphics[width=\hsize]{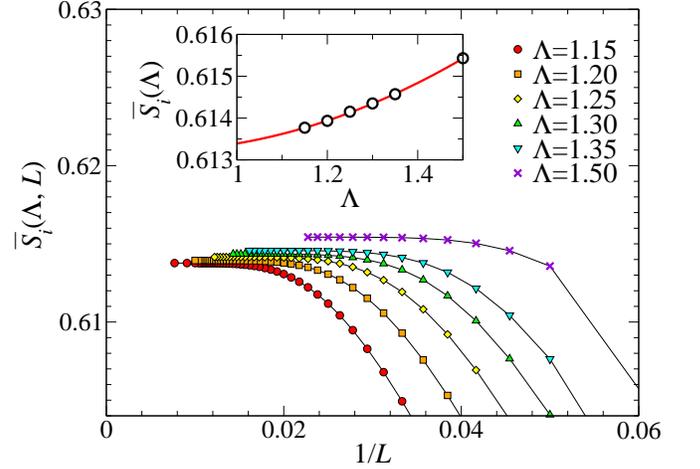}
\caption{(Color online) $L$ dependence of the total spin $\bar{S}_i(L,\Lambda)$ at impurity site 
for the single-impurity Anderson model on the honeycomb lattice at half-filling with $V = t$, $U=2t$,
and $\varepsilon = t$ (i.e., a particle-hole symmetric case). 
The logarithmic discretization scheme with various $\Lambda $ (indicated in the figure) is used. 
Here, we set $M = L$. 
The inset shows the extrapolated values of $\bar{S}_i(L,\Lambda)$ to $L \to \infty$, i.e., 
$\bar{S}_i(\Lambda)$, for six different $\Lambda$'s. 
A red line is a quadratic fit of data.
}
\label{fig:scaling_lambda}
\end{figure}

Next, we examine the $L$ and $\Lambda$ dependence of the total spin $\bar{S}_i$ at impurity site defined 
in Eq.~(\ref{eq:si}). In order to show more explicitly the $L$ and $\Lambda$ dependence, here we shall denote 
$\bar{S}_i$ as $\bar{S}_i(L,\Lambda)$ 
when it is calculated using the logarithmic discretization scheme.  
We find in Fig.~\ref{fig:scaling_lambda} that $\bar{S}_i(L,\Lambda)$ for a given $\Lambda$ is well converged 
when $L$ is sufficiently large. Note here that we set $M=L$ in Fig.~\ref{fig:scaling_lambda}. 
However, the extrapolated value of $\bar{S}_i(L,\Lambda)$ 
to $L\to\infty$, i.e.,
\begin{equation}
\bar{S}_i(\Lambda) = \lim_{L\to\infty} \bar{S}_i(L,\Lambda),
\label{eq:s_lambda}
\end{equation}
implying that $M\to\infty$ is also taken with a finite and fixed ratio of $M/L$,
exhibits slight but visible $\Lambda$ dependence. 
As shown in the inset of Fig.~\ref{fig:scaling_lambda}, we find that $\bar{S}_i(\Lambda)$ is almost linearly 
dependent on $\Lambda$ when $\Lambda $ is close to 1, and the difference between 
$\bar{S}_i(\Lambda)$ with $\Lambda=1.15$ and the extrapolated value to $\Lambda\to1^+$, i.e., 
\begin{equation}
\bar{S}_i^* = \lim_{\Lambda\to1^+} \bar{S}_i(\Lambda)
\label{eq:s*}
\end{equation}
is quite small $(\sim10^{-3})$. We thus conclude that $\bar{S}_i(\Lambda)$ with $\Lambda = 1.15$ 
can represent the value in the thermodynamic limit and 
the corresponding error is as small as $10^{-3}$. 
We should note here that the practical NRG calculations are typically performed with 
$\Lambda \ge 1.5$~\cite{osolin}.

\begin{figure}[htbp]
\includegraphics[width=\hsize]{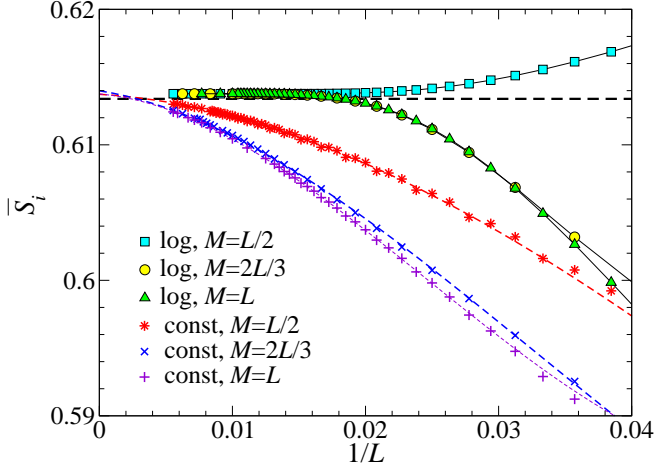}
\caption{(Color online) 
$L$ dependence of the total spin $\bar{S}_i$ at impurity site 
for the single-impurity Anderson model on the honeycomb lattice at half-filling with $V = t$, $U=2t$, 
and $\varepsilon = t$ (i.e., a particle-hole symmetric case). 
Here, we use both logarithmic and constant discretization schemes, denoted as  ``const'' and ``log,'' 
respectively, with $M=L/2$, $2L/3$, and $L$. 
We set $\Lambda=1.15$ for the logarithmic discretization scheme. 
Dashed lines along the symbols for the constant discretization scheme are fitting curvatures 
with cubic polynomials of $1/L$. 
A black dashed line indicates the extrapolated value $\bar{S}_i^*$ to $L\to\infty$ and 
$\Lambda\to1^+$ obtained 
in the inset of Fig.~\ref{fig:scaling_lambda}. 
}
\label{fig:scaling_L}
\end{figure}

Next, let us compare the results for the logarithmic discretization scheme and 
the constant discretization scheme. 
We indeed find in Fig.~\ref{fig:scaling_L} that, irrespectively of the discretization schemes, 
all results converge into a unique value 
in the limit of $L \to \infty$ within the error of $10^{-3}$. 
More precisely, the results in Fig.~\ref{fig:scaling_L} for the logarithmic and constant 
discretization schemes should converge to $\bar{S}_i(\Lambda)$ with $\Lambda=1.15$ and 
$\bar{S}_i^*$, respectively.
We also find in Fig.~\ref{fig:scaling_L} that the ratio $M/L$, which we set to be 1 in our calculations 
shown in the main text, dose not 
affect the converged value as long as 
$L$ and $M$ are sufficiently large. 
For the logarithmic discretization scheme, 
the difference for various $M/L$ is found to be as small as $10^{-9}$. 

\subsection{Phase boundary in the thermodynamic limit}

Here, we examine the phase boundary in the thermodynamic limit by 
explicitly taking the limits of $L\to\infty$ and $\Lambda\to1^+$ for the static quantities, 
i.e., the local density per spin $\bar n_{i\sigma}$ and the total spin $\bar S_i$ at the 
impurity site, calculated using the logarithmic discretization scheme. 
In order to show explicitly the $L$ and 
$\Lambda$ dependence of these quantities, here we adopt the convention used in Eqs.~(\ref{eq:s_lambda}) 
and (\ref{eq:s*}). Similarly, we take the limit of $L\to\infty$ for the local density per spin 
$\bar n_{i\sigma}(L,\Lambda)$ calculated for given $L$ and $\Lambda$, i.e., 
\begin{equation}
\bar{n}_{i\sigma}(\Lambda) = \lim_{L\to\infty} \bar{n}_{i\sigma}(L,\Lambda),
\end{equation}
and then take the limit of $\Lambda\to1^+$, i.e., 
\begin{equation}
\bar{n}_{i\sigma}^* = \lim_{\Lambda\to1^+} \bar{n}_{i\sigma}(\Lambda)
\end{equation}
to estimate the value in the thermodynamic limit. To obtain the well-converged and predictive values of
$\bar{S}_i(\Lambda)$ and $\bar{n}_{i\sigma}(\Lambda)$ within the residual error of $10^{-5}$, 
$L$ is required as large as 180 for $\Lambda = 1.15$.

\begin{figure}[htbp]
\includegraphics[width=\hsize]{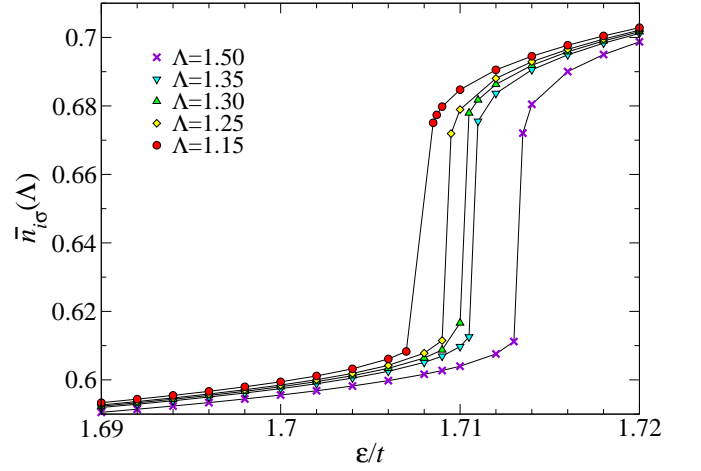}
\caption{(Color online) 
Local density per spin $\bar{n}_{i\sigma} (\Lambda)$ 
at impurity site for the single-impurity Anderson model on the honeycomb lattice 
at half-filling with $V=t$ and $U=2t$. The logarithmic discretization scheme 
with various $\Lambda$ (indicated in the figure) is used. 
}
\label{fig:scaling_n_param}
\end{figure}

\begin{figure}[htbp]
\includegraphics[width=\hsize]{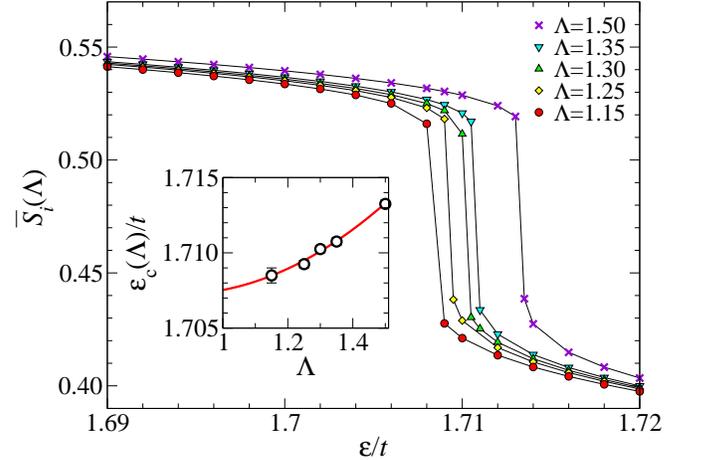}
\caption{(Color online) 
Total spin $\bar{S}_i (\Lambda)$ 
at impurity site for the single-impurity Anderson model on the honeycomb lattice 
at half-filling with $V=t$ and $U=2t$. The logarithmic discretization scheme 
with various $\Lambda$ (indicated in the figure) is used. 
The inset shows $\varepsilon_c(\Lambda)$ at which the abrupt change of $\bar{n}_{i\sigma} (\Lambda)$ and 
$S_i(\Lambda)$ occurs with varying $\varepsilon$. A red line is a fitting curve with a quadratic polynomial of 
$\Lambda$. 
}
\label{fig:scaling_param}
\end{figure}

Typical results around the phase boundary are shown in 
Figs.~\ref{fig:scaling_n_param} and \ref{fig:scaling_param}. 
We find in Figs.~\ref{fig:scaling_n_param} and \ref{fig:scaling_param} that 
$\bar{n}_{i\sigma}(\Lambda)$ and $\bar{S}_i(\Lambda)$ exhibit the abrupt changes exactly 
at the same $\varepsilon$, i.e., $\varepsilon_c(\Lambda)$, for each $\Lambda$, although 
$\varepsilon_c(\Lambda)$ itself depends clearly on $\Lambda$. The phase boundary in the thermodynamic 
limit is thus obtained by extrapolating $\varepsilon_c(\Lambda)$ to $\Lambda\to1^+$, i.e., 
\begin{equation}
\varepsilon_c^* = \lim_{\Lambda\to1^+} \varepsilon_c(\Lambda). 
\end{equation}
As shown in the inset of Fig.~\ref{fig:scaling_param}, we find that $\varepsilon_c^*=1.707t\pm 0.001t$ 
for $V=t$ and $U=2t$, which is in excellent agreement with $\varepsilon_c = 1.707t$--$1.708t$ 
estimated from the results for $L=128$ and $\Lambda = 1.15$ in the main text 
(Figs.~\ref{fig:ns} and \ref{fig:pd}). 
We should also note that $\varepsilon_c$ is very close to $\varepsilon_c(\Lambda) = 1.708$--$1.709t$
for $\Lambda = 1.15$. Therefore, we conclude that the typical error of the phase boundary $\varepsilon_c$ 
obtained in Figs.~\ref{fig:ns} and \ref{fig:pd} is as small as $0.002t$.

\subsection{Convergence behavior of dynamical quantities}

Finally, we discuss the convergence of the dynamical quantities $\chi_{s}(\omega)$ and 
$\chi_{c}(\omega)$ for a given broadening factor $\eta$. 
In order to make an accurate comparison, here we consider the noninteracting limit. 
In this limit, the spin excitation spectrum at the impurity site is given as 
\begin{eqnarray}
\chi_0 (\omega) = \frac{\eta}{2\pi} \sum_{e_k < \mu} \sum_{e_k^{\prime} > \mu} 
\frac{ \left| u_{i}^{(k)} u_{i}^{(k^{\prime})} \right|^2}{( \omega - e_k + e_{k^{\prime}})^2 + \eta^2}, 
\label{eq:chi0}
\end{eqnarray}
where $u_i^{(k)}$ is the impurity site component of the $k$th eigenstate of $\hat{H}_0^{\prime}$ 
in Eq.~(\ref{eq:tridiagonalform}) with its eigenvalue $e_k$. 
The charge excitation spectrum at the impurity site is expressed with the same form 
as in Eq.~(\ref{eq:chi0}) except for 
the additional factor 4, i.e., $\chi_c (\omega) = 4\chi_0(\omega)$, in the non-interacting limit. 
Therefore, we only consider $\chi_0 (\omega)$ below.

\begin{figure}[hbtp]
\includegraphics[width=\hsize]{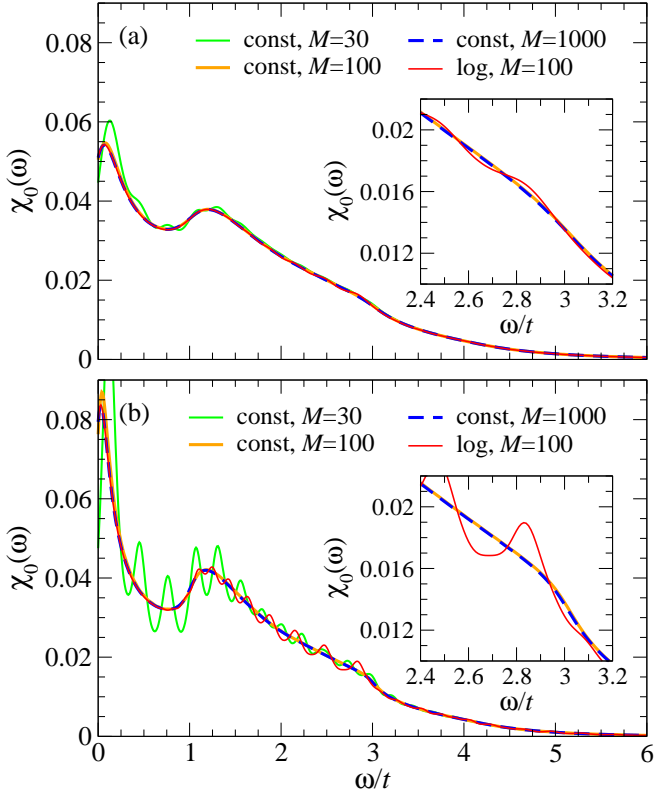}
\caption{(Color online) 
Spin and charge excitation spectrum $\chi_0 (\omega)$ at the impurity site in the noninteracting limit 
for the single-impurity Anderson model on the honeycomb lattice at half-filling with $V=t$ and $\varepsilon=0$. 
Here, we use both constant and logarithmic discretization schemes, denoted as ``const'' and ``log,'' respectively,  
with $M=L$ and broadening factors $\eta = 0.2t$ (a) and $0.1t$ (b). 
We set $\Lambda = 1.15$ for the logarithmic discretization scheme. 
The insets show the enlarged scale around $\omega \sim 2.4$--$3.2t$. For clarity, the results for the 
constant discretization scheme with $M=20$ and $40$ are not shown.
}
\label{fig:spec_conv_eta}
\end{figure}

As shown in Fig.~\ref{fig:spec_conv_eta}, we find that 
$\chi_0 (\omega)$ for smaller $L$ exhibits oscillating behavior when the constant discretization scheme is used. 
This is simply understood because the broadening factor $\eta$ is smaller than the level spacing 
of $e_k$ when $L$ is small. Therefore, the absence of such oscillating behavior 
for large enough $L$ is a hallmark of the convergence for a given $\eta$. Indeed, 
we find in Fig.~\ref{fig:spec_conv_eta} that $\chi_0 (\omega)$ for $L=100$ is well converged, 
as compared with that for $L=1000$, and the estimated error is as small as $10^{-3}$. 

We also notice in the insets of Fig.~\ref{fig:spec_conv_eta} that the similar oscillating behavior appears 
in the high-energy regions when the logarithmic discretization scheme is employed. 
This oscillating behavior is absent 
in the constant discretization scheme when the same $M$ is used. This is simply because, 
in the logarithmic discretization 
scheme, the energy mesh size is determined by $\Lambda$ and becomes wider in the higher-energy regions 
than in the lower-energy regions. In contrast, in the constant discretization scheme, the energy mesh size is 
constant for all energy regions and becomes smaller with increasing $M$ .

\begin{figure}[htbp]
\includegraphics[width=\hsize]{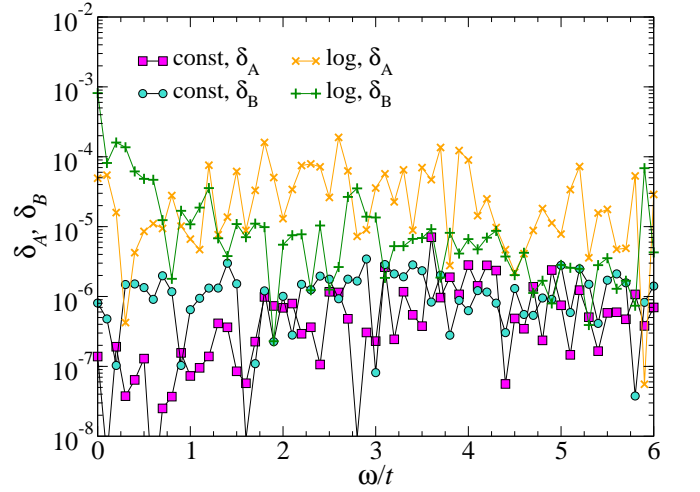}
\caption{(Color online) 
$\delta_A$ and $\delta_B$, defined in Eqs.~(\ref{eq:appliedoperror}) and (\ref{eq:specerror}), respectively, 
in the non-interacting limit for the single-impurity Anderson model on the honeycomb lattice at half-filling 
with $V=t$ and $\varepsilon = 0$. Here, we use both constant and logarithmic discretization schemes, 
denoted as ``const'' and ``log,'' respectively, with $M=L=100$ and a broadening factor $\eta = 0.2t$. 
We set $\Lambda = 1.15$ for the logarithmic discretization scheme. 
These results are obtained by performing six sweeps of the DMRG iteration with keeping $m_{\rm D}=1400$.
}
\label{fig:specprofile}
\end{figure}

Aside from the slow convergence problem in the high energy regions, there is a more serious technical issue 
for the logarithmic discretization scheme. 
In the dynamical DMRG calculation for, e.g., the spin excitation spectrum $\chi_s(\omega)$ at the impurity site, 
we have to construct the reduced density matrix of a mixed state composed of the ground state $|\psi_0 \rangle$ 
and the two excited states, 
\begin{eqnarray}
\left| A \right> = S_i^z \left| \psi_0 \right> 
\end{eqnarray}
and
\begin{eqnarray}
\left| B \right> = \frac{1}{ ( \omega - \mathcal{H} + E_0 )^2 + \eta^2 } S_i^z \left| \psi_0 \right>, 
\end{eqnarray}
for each $\omega$. 
Therefore, as compared with the ground state calculation, the convergence of the dynamical calculation is 
slower with respect to the number $m_{\rm D}$ of the reduced density matrix eigenstates kept in 
the DMRG calculation. 
This is an additional source of numerical errors and depends sensitively on the discretization schemes. 

Since 
\begin{eqnarray}
\left< A \left| A \right> \right. = \left< \psi_0 \right| S_i^z S_i^z \left| \psi_0 \right> = \frac{\bar{S}_i}{3} 
\end{eqnarray}
and
\begin{eqnarray}
\frac{\eta}{\pi} \left< B \right| \left[ ( \omega - \mathcal{H} + E_0 )^2 + \eta^2 \right] \left| B \right> = \chi_s (\omega), 
\end{eqnarray}
we can infer for each $\omega$ the errors in the two excited states $\left| A \right>$ and $\left| B \right>$ 
by evaluating 
\begin{eqnarray}
\delta_A = \left| \left< A \left| A \right> \right. - \frac{\bar{S}_i} {3} \right| 
\label{eq:appliedoperror}
\end{eqnarray}
and 
\begin{eqnarray}
\delta_B = \left| \frac{\eta}{\pi} \left< B \right| \left[ ( \omega - \mathcal{H} + E_0 )^2 + \eta^2 \right] \left| B \right> - \chi_s (\omega) \right| 
\nonumber \\
\label{eq:specerror}
\end{eqnarray}
in the non-interacting limit, where $\chi_s (\omega)$ is known exactly in Eq.~(\ref{eq:chi0}) and 
$\bar{S}_i = 1.5$ when the particle-hole symmetry is preserved. 

Figure~\ref{fig:specprofile} shows the results of $\delta_A$ and $\delta_B$ for the constant and 
logarithmic discretization schemes with keeping the same number $m_{\rm D}$ of the reduced 
density matrix eigenstates and the same tolerance for the optimization of $|\psi_0\rangle$ 
and $| B \rangle$~\cite{jeckelmann}. 
As shown in Fig.~\ref{fig:specprofile}, we find that both $\delta_A$ and $\delta_B$ for the logarithmic 
discretization scheme are larger than those for the constant discretization scheme. 
These results clearly show that the convergence of the excited states in the logarithmic discretization 
scheme is slow and thus a larger $m_{\rm D}$ is required to reach 
the same convergence as in the constant discretization scheme, implying that the logarithmic discretization 
scheme demands more computational cost. 
Therefore, we employ the constant discretization scheme to calculate the full excitation spectra.


\begin{thebibliography}{99}
\bibitem{note7} 
To be more precise, these three lowest entanglement levels are still quasi degenerate at 
$\varepsilon=1.707t$ as 
$\xi_1=\xi_2=1.429$ and $\xi_3=1.609$. When we increase $\varepsilon$ slightly to $1.708t$, 
we find that $\xi_1=1.292$ and $\xi_2=\xi_3=1.584$.
Further fine tuning of $\varepsilon$ is required to reach at the exact three fold degenerate point, 
corresponding to the VF point at the phase boundary.
\bibitem{note9}
When $l_{\rm L}$ is even and $L$ is finite, $|\bar\psi_1\rangle\sim|\phi_{\uparrow}\rangle_{\rm L}\otimes|0\rangle_{\rm R}$ 
and $|\bar\psi_2\rangle\sim|\phi_\downarrow\rangle_{\rm L}\otimes|0\rangle_{\rm R}$ are entangled, where 
$|\phi_{\uparrow(\downarrow)}\rangle_{\rm L}\sim|\uparrow(\downarrow)\rangle_i\otimes|\downarrow(\uparrow)\rangle_{\rm \bar L}$ and $|s\rangle_{\rm \bar L}$ with $s=\,\uparrow,\,\downarrow,\,0$ represents the $z$ component of total spin ($Sz=1/2,-1/2,0)$ in the left block 
excluding the impurity site. Therefore, the ground state is represented approximately as $|\psi_0\rangle\sim\frac{1}{\sqrt{2}}\left(|\bar\psi_1\rangle-|\bar\psi_2\rangle\right)\sim\frac{1}{\sqrt{2}}\left(|\phi_{\uparrow}\rangle_{\rm L}-|\phi_{\downarrow}\rangle_{\rm L}\right)\otimes|0\rangle_{\rm R}$. We can now easily show that the lowest $\xi_k$ 
for $|\psi_0\rangle$ is singlet.
\bibitem{note8} 
In the ASC phase for $\varepsilon > U/2$, 
we find that the lowest entanglement level is non degenerate for both even and odd $l_L$. 
This is because the total number of electrons in the left block is $l_L$ for $l_L$ even and 
$l_L+1$ for $l_L$ odd.
\end{thebibliography}
\end{document}